\documentclass[final,5p,times,twocolumn,authoryear]{elsarticle}

\usepackage{graphicx}      
\usepackage{natbib}        
\usepackage{hyperref}
\usepackage{orcidlink}
\usepackage{enumitem}
\usepackage{pbox}
\usepackage{booktabs}      
\usepackage{array}
\usepackage{amssymb}
\usepackage{float}
\usepackage{subfig}
\usepackage{subcaption}
\usepackage{caption}
\usepackage{amsmath}
\usepackage{multicol}
\usepackage{multirow}
\usepackage{tikz}
\usepackage{forest}
\usepackage{comment}
\usepackage{balance}
\makeatletter
\newcommand{\numberedsubfigures}{%
	\renewcommand{\thesubfigure}{\arabic{subfigure}}%
	\renewcommand{\p@subfigure}{\thefigure.}%
}
\renewcommand{\paragraph}[1]{%
	\@startsection{paragraph}{4}{\parindent}%
	{0pt}%
	{-1em}%
	{\normalfont\normalsize}%
	{\textit{#1.}}%
}
\makeatother

\definecolor{folderbg}{RGB}{124,166,198}
\definecolor{folderborder}{RGB}{110,144,169}
\definecolor{calgrey}{RGB}{110,110,110}
\definecolor{ccblue}{RGB}{0,102,204}
\definecolor{dynred}{RGB}{192,0,0}
\def\NextBMSFolderSize{4pt}
\tikzset{
	nextbmsfolder/.pic={
		\filldraw[draw=folderborder,top color=folderbg!50,bottom color=folderbg]
			(-1.05*\NextBMSFolderSize,0.2\NextBMSFolderSize+5pt) rectangle ++(.75*\NextBMSFolderSize,-0.2\NextBMSFolderSize-5pt);
		\filldraw[draw=folderborder,top color=folderbg!50,bottom color=folderbg]
			(-1.15*\NextBMSFolderSize,-\NextBMSFolderSize) rectangle (1.15*\NextBMSFolderSize,\NextBMSFolderSize);
	}
}
\forestset{
	nextbms folder tree/.style={
		for tree={
			font=\ttfamily\scriptsize,
			grow'=0,
			child anchor=west,
			parent anchor=south,
			anchor=west,
			calign=first,
			inner xsep=7pt,
			inner ysep=1pt,
			s sep=1pt,
			edge path={
				\noexpand\path [draw, \forestoption{edge}]
				(!u.south west) +(7.5pt,0) |- (.child anchor) pic {nextbmsfolder} \forestoption{edge label};
			},
			before typesetting nodes={
				if n=1
					{insert before={[,phantom]}}
					{}
			},
			fit=band,
			before computing xy={l=13pt},
		}
	}
}

\usepackage{cleveref}
\Crefformat{figure}{#2Fig.~#1#3}
\Crefmultiformat{figure}{Figs.~#2#1#3}{ and~#2#1#3}{, #2#1#3}{ and~#2#1#3}
\Crefformat{equation}{#2Eq.~#1#3}
\Crefformat{section}{#2Section~#1#3}
\Crefmultiformat{equation}{Eqs.~#2#1#3}{ and~#2#1#3}{, #2#1#3}{ and~#2#1#3}
\Crefformat{table}{#2Table~#1#3}

\usepackage{todonotes}
\newcounter{todocounter}
\newcounter{ga}

\newcommand{\rev}[1]{#1}

\usepackage[acronym]{glossaries} 
\newacronym{evs}{EVs}{Electric Vehicles}
\newacronym{soc}{SoC}{State-of-Charge}
\newacronym{soh}{SoH}{State-of-Health}
\newacronym{dod}{DoD}{Depth-of-Discharge}
\newacronym{hm}{HM}{Hybrid Model}
\newacronym{pb}{PB}{Physics Based}
\newacronym{ml}{ML}{Machine Learning}
\newacronym{cc}{CC}{Constant-Current}
\newacronym{cv}{CV}{Constant-Voltage}
\newacronym{gitt}{GITT}{Galvanostatic Intermittent Titration-Technique}
\newacronym{hppc}{HPPC}{Hybrid Power-Pulse Characterisation }
\newacronym{rul}{RUL}{Remaining Useful Life}
\newacronym{eis}{EIS}{Electrochemical Impedance Spectroscopy}
\newacronym{sem}{SEM}{Scanning Electron Microscope}
\newacronym{edx}{EDX}{Energy-Dispersive X-ray Spectroscopy}
\newacronym{bol}{BoL}{Beginning-of-Life}
\newacronym{eol}{EoL}{End-of-Life}
\newacronym{mol}{MoL}{Middle-of-Life}
\newacronym{rpt}{RPT}{Reference Performance Tests}
\newacronym{ocp}{OCP}{Open-Circuit Potential}
\newacronym{ocv}{OCV}{Open-Circuit Voltage}
\newacronym{lvc}{LVC}{Lower Voltage Cut-off}
\newacronym{sei}{SEI}{Solid Electrolyte Interphase}
\par
\journal{Journal of Power Sources}
\par
\begin{document}
\makeatletter
\def\lb@subsubsection{\thesubsubsection.\hskip 0.5em}
\setcounter{secnumdepth}{3}
\makeatother
\begin{frontmatter}
\par
\par
\title{Characterization, ageing and post-mortem dataset of commercial 58Ah prismatic NMC cells} 
\par
\author[TNO_label]{R\'obinson Medina\,\orcidlink{0009-0001-2214-6153}}
\author[TNO_label]{Alessio A. Lodge\,\orcidlink{0009-0001-5478-0183}}
\author[TNO_label]{Feye S.J. Hoekstra\,\orcidlink{0000-0002-1471-7489}}
\author[TNO_label,TUe_label]{Steven Wilkins\,\orcidlink{0000-0001-9498-2321}}
\author[AIT_label]{Damian Cupid\,\orcidlink{0000-0002-2622-9060}}
\author[AIT_label]{Akzhan Bekzhanov\,\orcidlink{0000-0001-5842-1383}}
\author[AIT_label]{Dominik Dvorak\,\orcidlink{0000-0003-2856-5882}}
\author[AIT_label]{Hansjörg Kapeller\,\orcidlink{0000-0003-3626-713X}}
\author[NIC_label]{Gregor Kapun\,\orcidlink{0000-0003-0007-4678}}
\author[NIC_label,UL_label]{Jože Moškon\,\orcidlink{0000-0002-8223-0031}}
\author[NIC_label,UL_label]{Blaž Tratnik\,\orcidlink{0009-0003-8166-8941}}
\author[UL_label]{Igor Mele\,\orcidlink{0000-0002-9071-7325}}
\author[UL_label]{Tomaž Katrašnik\,\orcidlink{0000-0001-6954-4936}}
\par
\affiliation[TNO_label]{organization={TNO},
            addressline={Automotive Campus 30}, 
            city={Helmond},
            postcode={5708 JZ}, 
            country={The Netherlands}}
\par
\affiliation[TUe_label]{organization={Eindhoven University of Technology},
            addressline={Groene Loper 3}, 
            city={Eindhoven},
            postcode={5612 AE}, 
            country={The Netherlands}}
\par
\affiliation[AIT_label]{organization={Austrian Institute of Technology GmbH},
            addressline={Giefinggasse 2}, 
            city={Vienna},
            postcode={1210}, 
            country={Austria}}
\par
\affiliation[NIC_label]{organization={National Institute of Chemistry},
            addressline={Hajdrihova 19}, 
            city={Ljubljana},
            postcode={1000}, 
            country={Slovenia}}
\par
\affiliation[UL_label]{organization={University of Ljubljana, Faculty of Mechanical Engineering, LICeM},
            addressline={Aškerceva 6}, 
            city={Ljubljana},
            postcode={1000}, 
            country={Slovenia}}
\par
\begin{abstract}                
\par
This paper provides an extensive dataset for commercial 58 Ah prismatic NMC cells. The dataset comprises three types of data: characterization, ageing, and cell teardown.
Characterization data include tests such as GITT, HPPC, and EIS, among others.
Ageing data were obtained from 41 cells subjected to 38 distinct ageing conditions, including application-surrogate drive cycles.
Recurring reference performance tests and EIS measurements track the evolution of each cell.
Teardown data were obtained from pristine and aged (post-mortem) cells. They include material analyses and half-cell measurements.
This single dataset supports analytics, model development (both machine-learning and physics-based), benchmarking, and techno-economic analysis.
Several example analyses are carried out to show the usability of the data. For instance, the paper shows how lithium stripping can be detected from the voltage response of a short discharge applied after each charge.
The data and the processing and analysis scripts are openly provided alongside this paper.
\par
\end{abstract}
\par
\begin{graphicalabstract}
\centering
\includegraphics[width=\textwidth]{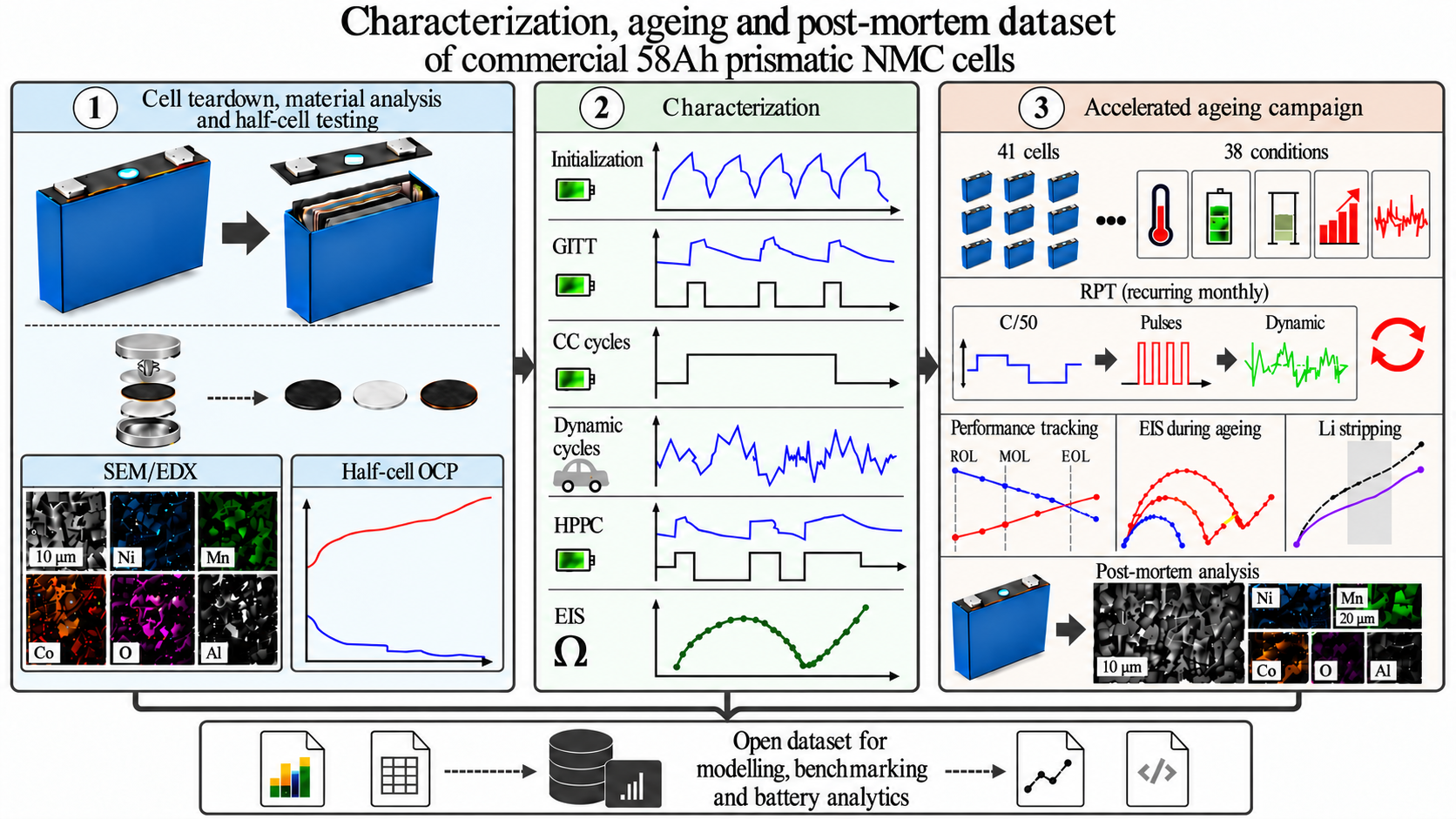}
\end{graphicalabstract}
\par
\begin{keyword}
Battery data \sep Ageing dataset \sep Lithium-ion batteries \sep Cell teardown data \sep Post-mortem analysis
\par
\end{keyword}
\par
\end{frontmatter}
\section{Introduction}\label{sec:introduction}
Lithium-ion batteries are the fastest-growing energy-storage technology: they provide sufficient energy density for stationary and mobile applications and already support the decarbonization of transport and industrial energy storage \citep{iea2026globalenergyreview}. Batteries in \gls{evs} provide zero local emissions \citep{kelly2023cradle}, and utility-scale batteries provide grid-scale flexibility and grid services \citep{iea2024batteries}. When installed behind the meter alongside rooftop solar, coupled to large industrial loads, or co-located with large generation sites (utility-scale solar or wind farms), batteries can reduce the need for fossil-fuel-based energy by reducing curtailment, improving self-consumption, and increasing the penetration of renewables \citep{prakash2022review}.
\par
All modern lithium-ion battery systems in \gls{evs} or stationary applications require at least one algorithm or model to estimate \gls{soc}, which cannot be measured directly \citep{plett2015battery}. As the technology matures, advanced algorithms contribute to improving battery usage and performance. Applications include state estimation \citep{lopetegi2024new,reniers2021unlocking}, design support \citep{kench2024li}, system sizing \citep{grover2018optimal, zhang2023optimal}, fast-charging \citep{wassiliadis2023model} and advanced analytics (e.g., State-of-Health, State-of-Safety (SoS), and Remaining Useful Life \citep{li2021one, li2025fast, roy2024forecasting, wang2023large, wang2024unlocking}). All models, whether empirical \citep{plett2015battery}, Physics-Based \citep{plett2024battery}, Machine Learning \citep{panahi2025fast}, or Hybrid Models \citep{borah2024synergizing}, require battery information and data for their applications \citep{sheikh2026pybatteryid, chen2020development, hoekstra2023rapid, brosa2022continuum}. The amount and type of information required vary by model. Despite the wide adoption of battery systems, battery manufacturers and operators generally disclose only limited information, while detailed data, such as time-series measurements or characterization tests, are not provided.
\par
Several commercial providers and online platforms have emerged to address this gap, including About:Energy \citep{aboutenergy}, BatteryDesign.net \citep{batterydesign}, and Batemo \citep{batemo}. In this work, three categories of battery data are distinguished: (i) characterization data, i.e., electrical and other measurements on pristine cells \citep[e.g.,][]{catenaro2021experimental,fasolato2025dataset,khan2025high}; (ii) ageing data, i.e., measurements on cells cycled or stored over extended periods to capture changes in performance \citep[e.g.,][]{SCHMITT2017183,frie2024analysis,kirkaldy2024lithium,stroebl2024multi,luh2024comprehensive,khan2025calendar,pozzato2022lithium,moy2024second,Cui_2025}; and (iii) teardown data, i.e., electrochemical, chemical, visual, and physical analyses at the electrode level, including half-cell testing of harvested electrodes \citep[e.g.,][]{ank2023lithium, stock2023cell, gunter2022state}. Many datasets cover at least one of these types of data. However, to the best of our knowledge, no open dataset combines characterization, ageing, and teardown data from the same cell, together with the processing scripts needed to use them. Therefore, models built for a specific cell often combine elements from different sources while still lacking some of the information needed from a single cell in the literature.
\par
To fill this gap, we present an open dataset composed of characterization, ageing, and teardown data acquired from commercial 58Ah prismatic NMC cells with a graphite anode and a monocrystalline NMC 532 cathode. The characterization part of the data was measured on three cells and comprises initialization, \gls{gitt}, \gls{cc}, dynamic-cycle, \gls{hppc}, and \gls{eis} tests. The ageing part of the data covers 41 cells under 38 distinct electrical and thermal conditions. The teardown part of the data covers pristine and aged (post-mortem) cells. They include electrode dimensions, half-cell \gls{ocp} measurements, \gls{sem} imaging, and \gls{edx} elemental mapping of the electrodes. The dataset is accompanied by processing and plotting scripts. In addition, the paper presents example analyses of each type of data to illustrate its usability; these examples are not intended as an exhaustive analysis.
\par
This study was conducted across laboratories of three institutions: the Netherlands Organisation for Applied Scientific Research (TNO), AIT Austrian Institute of Technology GmbH (AIT), and the University of Ljubljana (UL). TNO performed the characterization; TNO and AIT shared the ageing experiments equally; UL and the National Institute of Chemistry performed pristine teardown and post-mortem analysis.
\par
\section{CALB Cell}\vspace{-0.2cm}\label{sec:comparison_methodology}
The cell selected and characterized in this study is a prismatic CALB cell with a nominal capacity of 58Ah, marketed as \textit{generation 3b}. The manufacturer's datasheet specifies a maximum discharge rate of 1.0C and a maximum charge rate of 1.2C (\Cref{tab:Cell_datasheet}). Based on the \gls{edx} analysis (\Cref{subsec:electrode_composition}), the cell was found to have an NMC 532 (55:30:15) cathode and a graphite anode. The investigated cells were selected as commercially available off-the-shelf components, ensuring high reproducibility and comparability of the results.
Furthermore, they represent realistic cell technology already available in mobile and stationary systems, while providing a favourable trade-off between cost and performance.
\begin{table}[h!]
\centering
\caption{Specifications from the manufacturer's datasheet}
\begin{tabular}{@{}ll@{}}
\toprule
\textbf{Property} & \textbf{Value} \\
\midrule
Chemistry & NMC \\
Form factor & Prismatic \\
Weight & 926g \\
Dimensions & \begin{tabular}[t]{@{}ll@{}}
    Thickness: 27mm \\
    Height: 106mm \\
    Length: 148mm \\\end{tabular} \\
Operating temperature &
\begin{tabular}[t]{@{}ll@{}}
    -20°C to 55°C (charging) \\
    -30°C to 55°C (discharging)\end{tabular} \\
Work voltage & 
  \begin{tabular}[t]{@{}ll@{}}
    2.75V to 4.35V ($T \geq 0^\circ\mathrm{C}$) \\
    2.2V to 4.35V (-30°C to 0°C)\end{tabular} \\
Rated voltage & 3.67V \\
Rated capacity & 58Ah \\
Max. charging rate & 1.2C (0°C to 45°C) \\
Max. discharging rate & 1.0C (20°C to 40°C) \\
Internal resistance & $0.60\mathrm{m}\Omega$ to $0.80\mathrm{m}\Omega$ \\
Charging time  & Approx. 4h \\
\gls{soc} Window & 5\% to 97\% \\
Cycle life & $\geq 2000$ cycles \\
\bottomrule
\label{tab:Cell_datasheet}
\end{tabular}
\end{table}
\par
A total of 100 pristine cells were acquired and divided into subsets for the different types of experiments. In general, each part of the experimental campaign used ``fresh cells'', i.e., cells that had not undergone previous cycling during this study, except where noted otherwise.
Incoming cells were screened for weight, dimensions, cell voltage, and physical condition. Eighteen dented or otherwise damaged cells were excluded from testing.
\par
\begin{figure*}[btp]
    \centering
    \numberedsubfigures
    \subfloat[Cell fixture.]{%
        \begin{minipage}[t]{0.31\textwidth}
            \includegraphics[width=\linewidth,height=0.24\textheight]{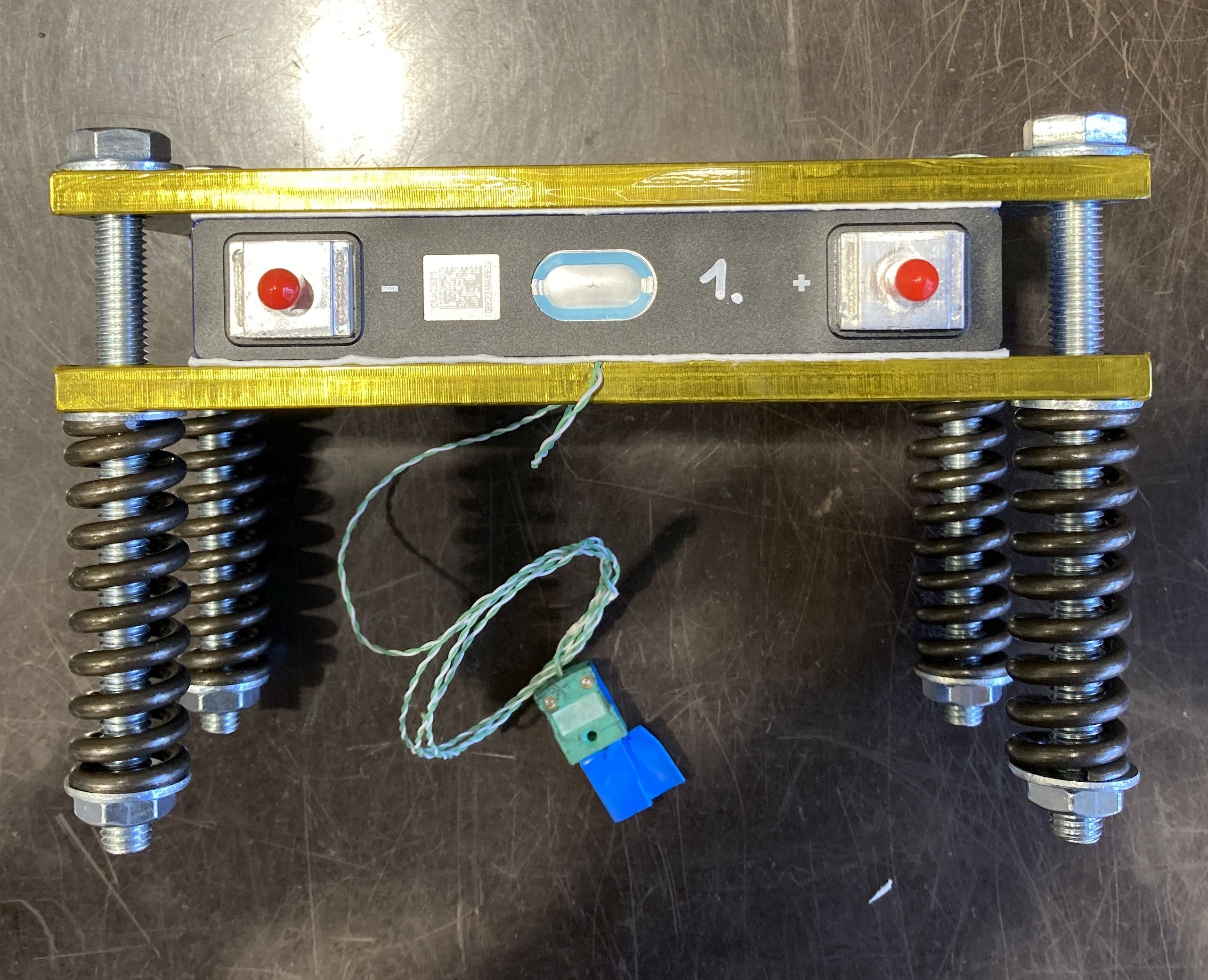}%
        \end{minipage}%
        \label{fig:cell_fixture}%
    }
    \hfill
    \subfloat[CALB 58Ah cell during teardown.]{%
        \begin{minipage}[t]{0.31\textwidth}
            \includegraphics[width=\linewidth,height=0.24\textheight]{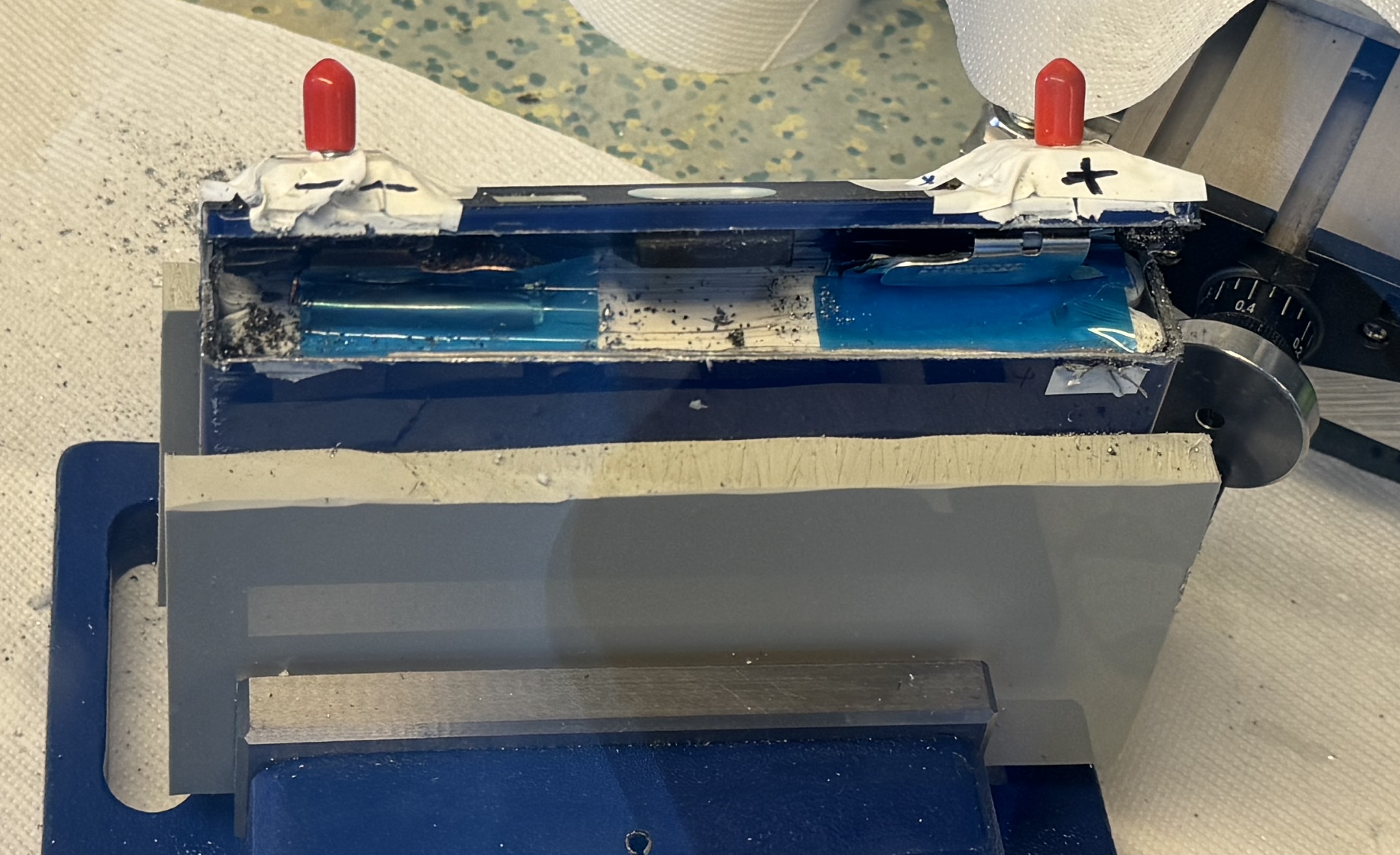}%
        \end{minipage}%
        \label{fig:cell_choice}%
    }
    \hfill
    \subfloat[Pristine-cell disassembly stages.]{%
        \begin{minipage}[t]{0.31\textwidth}
            \includegraphics[width=\linewidth,height=0.24\textheight]{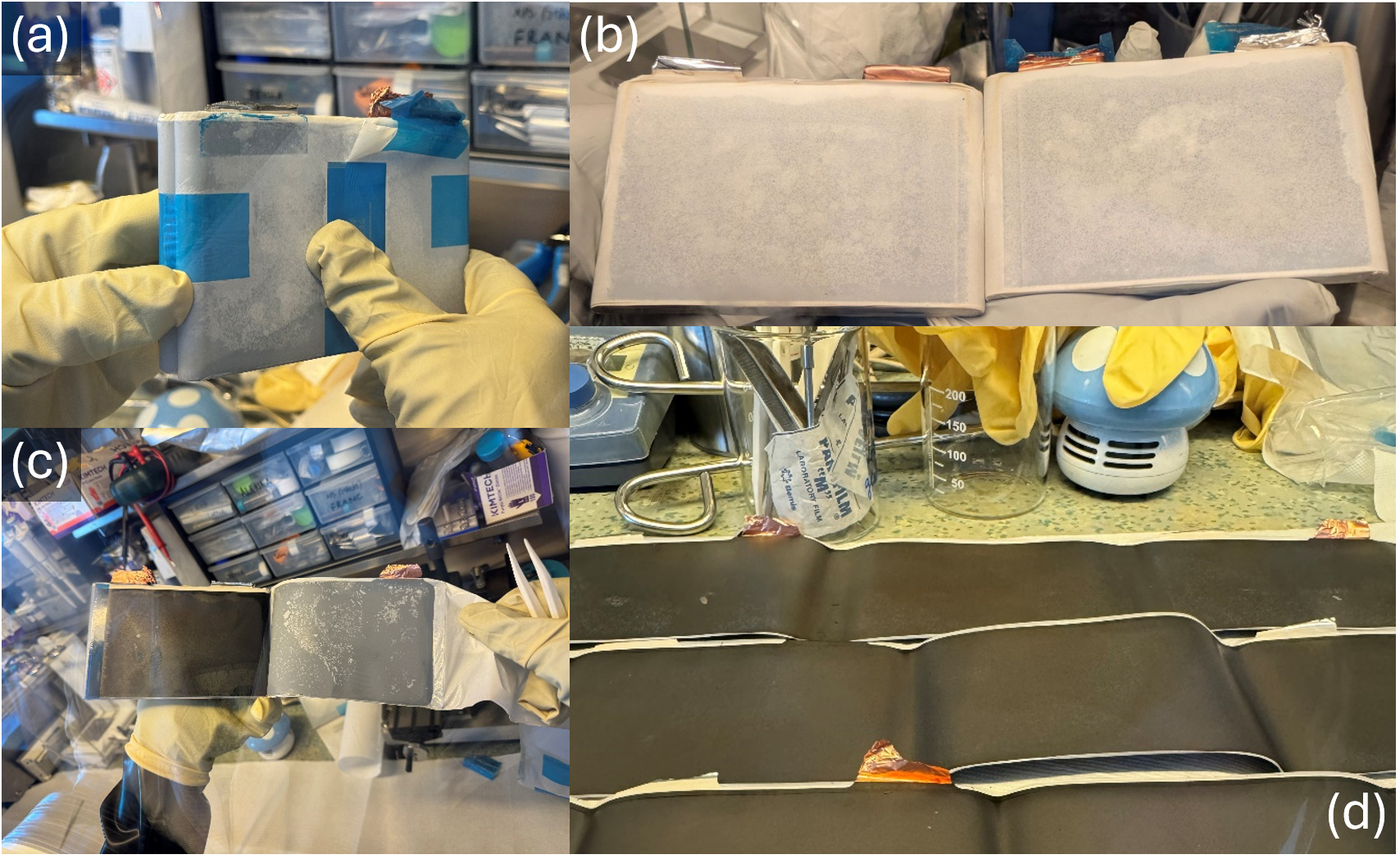}%
        \end{minipage}%
        \label{fig:cell_teardown_stages}%
    }
    \caption{Cell and teardown overview.}
    \label{fig:cell_overview}
\end{figure*}

Prismatic cells are known to have pressure-dependent performance, particularly when aged \citep{ding2025}.
For this reason, maintaining consistently applied pressure is an important feature of the study.
A fixture was therefore devised for all electrically tested cells (\Cref{fig:cell_fixture}).
Cells were positioned between two parallel steel plates, with each of the two largest faces of the cell in contact with one plate.
One plate has pass-through M6 bore holes, and the other has threaded bore holes, aligned so that the plates mate.
Threaded rods, washers, and nuts hold the plates together.
Four identical springs per cell, compressed by the same displacement, provide constant pressure.
The applied displacement was 2.57mm from an initial spring length of 70mm; at a spring rate of 95.508N/mm, this results in a force of \rev{245.5N per spring} and a total applied force (four springs) of \rev{982N (100kgf)} as specified by the cell manufacturer\rev{, who defines the cell thickness at 100$\pm$10kgf and 70\% \gls{soc}}.
To measure the temperature of each cell, a thermocouple was positioned in the middle of the bottom main face along the central slot (\Cref{fig:cell_fixture}).\par
\par
\par
\section{Methods}\label{Sec:Methods}
The experimental work described in this manuscript and the accompanying data comprise three main elements: (1) cell teardown, material analysis, and half-cell testing; (2) cell-level electrical characterisation; and (3) cell-level accelerated ageing testing. The campaign aims to produce a dataset for training and testing battery models used in design and operation, such as Remaining Useful Life algorithms. The dataset can also support benchmarking, analytics, and techno-economic analysis.
\subsection{Cell teardown }\label{subsec:teardown}
The purpose of the cell teardown is to reveal the cell's jelly rolls and expose the electrodes for further analysis. We distinguish two types of analysis:
\subsubsection{Electron microscopy and EDX analysis}
This included geometrical and half-cell measurements, optical and \gls{sem} imaging of the electrodes, and \gls{edx} elemental analysis.
Both pristine and aged cells were analysed. Before teardown, pristine cells were discharged to the \gls{lvc} using a CC-CV procedure to minimise their energy content. Aged cells were discharged below the \gls{lvc} to 1.5V and held there using a CC-CV procedure for at least 2days.\par
\par
Once discharged, the cells were placed in an MBraun glovebox under an inert atmosphere, where all subsequent operations up to half-cell harvesting were performed.
The cell housing was mechanically removed, taking care to avoid contact with the jelly roll or tabs. The electrolyte was first extracted and subjected to further analysis to determine electrolyte components. Afterwards, the jelly rolls were extracted by pulling them out with clippers. The cell during teardown and the different disassembly stages, including both jelly rolls and fully rolled-out electrodes, are shown in \Cref{fig:cell_choice,fig:cell_teardown_stages}.
The geometric properties of both electrodes, the separator, and the current collectors were analysed by unravelling the jelly roll in the glovebox. The width and length of the electrode strips were then measured with rulers, while thicknesses were determined with a micrometer (Tactix).
Circular 15mm electrode punches were used to prepare samples for further electrochemical characterisation and \gls{sem} imaging.\par
The \gls{sem} imaging was performed on a field-emission scanning electron microscope (Thermo Fisher Scientific/FEI Nova NanoSEM 450) equipped with an Everhart–Thornley secondary-electron (ETD) detector and a segmented in-lens detector (T1/T2/T3 with scintillator), operated at 1–2kV over 80×–50 000×.
Secondary-electron images were acquired using both the ETD and the in-lens segments (A+B).
Further analysis was carried out by visual inspection.
\par
\begin{table*}[!t]
  \centering
  \caption{Overview of the ageing campaign (41 cells). Rows place each cell by
  the degradation factor it isolates and columns give the test temperature.
  C-rates are listed as charge/discharge in C, and the
  dynamic profiles are stationary (stat.) or automotive (auto.). For the
  constant-current cells, \gls{dod} = 100\% and average \gls{soc} = 50\% unless shown
  otherwise. $^{a}$\,Cell\_12 is the 25°C baseline, repeated as Cell\_56,
  Cell\_89 and Cell\_93 (started after middle-of-life) to gauge consistency.
  $^{\ddag}$\,Cell\_66, Cell\_71, Cell\_72 and Cell\_74 at 0°C were
  compromised during the middle-of-life \gls{eis} and subsequently removed.}
  \label{tab:ageing_matrix}
  \footnotesize
  \setlength{\tabcolsep}{4pt}
  \renewcommand{\arraystretch}{1.15}
  \begin{tabular}{@{}>{\raggedright\arraybackslash}p{0.155\textwidth}
                     >{\raggedright\arraybackslash}p{0.186\textwidth}
                     >{\raggedright\arraybackslash}p{0.186\textwidth}
                     >{\raggedright\arraybackslash}p{0.186\textwidth}
                     >{\raggedright\arraybackslash}p{0.186\textwidth}@{}}
    \toprule
    \textit{Factor isolated}
      & \centering\textbf{0°C}
      & \centering\textbf{25°C}
      & \centering\textbf{45°C}
      & \centering\arraybackslash\textbf{0--45°C} \\
      & \centering TNO
      & \centering TNO
      & \centering AIT
      & \centering\arraybackslash AIT \\
    \midrule
\par
\textcolor{calgrey}{\textbf{Calendar}}\newline\textit{SoC}
      & Cell\_57 (100\%)
      & Cell\_11 (100\%)
      & Cell\_45 (10\%)\newline Cell\_26 (50\%)\newline Cell\_28 (100\%)
      & \\
    \midrule
\par
\textcolor{ccblue}{\textbf{Baseline}}\newline\textit{0.5C/0.5C}
      & Cell\_60
      & \textbf{Cell\_12$^{a}$, 56, 89, 93}
      & Cell\_29
      & Cell\_49 \\
    \addlinespace
\par
\textcolor{ccblue}{\textbf{C-rate}}\newline\textit{ch/dis (C)}
      & Cell\_63 (0.25/0.5)\newline Cell\_66$^{\ddag}$ (0.75/0.5)\newline
        Cell\_68 (1.0/0.5)\newline Cell\_71$^{\ddag}$\newline
        (0.25, 0.5, 0.75, 1.0~/~0.5)
      & Cell\_23 (1.0/0.5)\newline Cell\_34 (1.5/0.5)\newline
        Cell\_35 (2.0/0.5)\newline Cell\_43 (0.5/1.5)
      & Cell\_9 (1.0/0.5)\newline Cell\_8 (0.5/1.0)\newline
        Cell\_47 (1.0/1.0)
      & \\
    \addlinespace
\par
\textcolor{ccblue}{\textbf{DoD}}\newline\textit{0.5C/0.5C}
      &
      & Cell\_16 (10\%)\newline Cell\_30 (40\%)\newline Cell\_27 (70\%)
      &
      & \\
    \addlinespace
\par
\textcolor{ccblue}{\textbf{Average SoC}}\newline\textit{50\% \gls{dod}, 0.5C/0.5C}
      &
      &
      & Cell\_40 (75\%)\newline Cell\_1 (25\%)\newline Cell\_3 (50\%)
      & \\
    \addlinespace
\par
\textcolor{ccblue}{\textbf{High voltage}}\newline\textit{1.0C/0.5C}
      &
      &
      & Cell\_5 (4.45V)\newline Cell\_22 (4.45V, CV)
      & \\
    \midrule
\par
\textcolor{dynred}{\textbf{Dynamic}}\newline\textit{profile}
      & Cell\_74$^{\ddag}$ (stat.)\newline Cell\_72$^{\ddag}$ (auto.)
      & Cell\_46 (stat.)\newline Cell\_42 (auto.)
      & Cell\_17 (stat.)\newline Cell\_25 (auto.)
      & Cell\_50 (stat.)\newline Cell\_53 (auto.)\newline
        Cell\_64 (mix)\newline \hspace*{0.6em}auto 1-4, stat 5-6,\newline
        \hspace*{0.6em}auto 7-8, stat 9-12~mo\newline
        -Cell\_70 (mix)\newline \hspace*{0.6em}auto 1-6, stat 7-12~mo \\
    \bottomrule
  \end{tabular}
\end{table*}
\subsubsection{Half-cell testing}\label{subsec:half-cell testing}
Half-cells were prepared from 15mm circular punches of double-coated electrode. The coating on one side was removed by mechanical scraping with a ceramic knife and cleaning with NMP solvent at room temperature.
NMC 532 cathode and graphite anode samples were assembled as 2032 coin cells, each tested in a half-cell configuration against a lithium chip.
During assembly, a polypropylene separator was used, and each cell was wetted with 500µL of LP30 electrolyte (1M LiPF$_6$ in EC:DMC, 1:1 v/v). The separator and electrolyte did not exactly match those of the full cell.
After assembly, the half-cells were cycled outside the glovebox to obtain \gls{ocp} measurements.\par
\subsection{Electrical characterisation}\label{subsec:eletrical_char}
Electrical characterisation tests were performed to parametrise empirical, data-driven, and electrochemical models. The procedures were designed so that different model elements, e.g., \gls{ocv}, and their dependencies could be distinguished and modelled separately. To assess the consistency of the acquired cells, all characterisation procedures were performed on three cells: Cell\_4, Cell\_6, and Cell\_15. The cells were placed in a Weiss Technik 81002 climate chamber and cycled using a Chroma 7100 Cell Cycler (100A per channel, $\pm$5V).
For \gls{eis} at 25°C, the cells remained in the climate chamber. The Chroma Cycler channels were disconnected, and the Ivium 5000 \gls{eis} channels were connected through the climate-chamber routing. The applied characterisation tests are shown in full in \Cref{fig:CharacterizationHPPC}, and their individual elements are outlined below.
\par
\paragraph{\rev{Initialization}} Five consecutive \gls{cc} \gls{cv} C/4 charges and C/4 \gls{cc} discharges ensured uniform starting conditions for all cells.
\par
\paragraph{\gls{gitt}} Charge and discharge \gls{gitt} tests were performed to extract \gls{ocv} curves. Each test consists of 25 rest points per current direction, with transitions between points using 0.2C pulses of fixed duration followed by an adaptive rest period (the next pulse starts when relaxation falls below 1mV/hour) at 25°C. At every fifth point, after the relaxation condition is met, a temperature sweep is performed (as shown in the bottom subplot of \Cref{fig:CharacterizationHPPC}) to map the temperature dependence of \gls{ocv}.
\par
\paragraph{\gls{cc}} Cells were charged and discharged at C/20 at 25, 45, 5, and $-5^\circ\mathrm{C}$; they were then kept at 25°C and cycled at 0.1C, 0.2C, 0.5C, and 1.0C.
\par
\paragraph{Dynamic Cycles} Two distinct drive cycles were applied at $-5$, 5, 25, and 45°C, and 0, 15, 25, and 35°C, in both discharge and charge directions.
\par
\paragraph{\gls{hppc}} Short CC charge and discharge pulses were performed at various C-rates, temperatures, and \gls{soc} levels. For this study, pulses at 0.1C, 0.2C, 0.5C, 1.0C, and 1.5C were applied consecutively, each lasting 30s with 10min of rest between pulses. The sequence was measured at 25, 5, and 45°C and at \gls{soc} levels of 100\%, 80\%, 60\%, 40\%, 20\%, and 0\%.
\paragraph{\gls{eis}} \gls{eis} tests were executed in galvanostatic mode at 5.8A, ranging from 100kHz to 10mHz, at multiple \gls{soc} points (90\%, 50\%, 10\%, defined by voltage) and temperatures of 0, 5, 25 and 45°C. To reach the next \gls{soc} point, the cells were heated or cooled to a reference temperature of 25°C in the climate chamber, held there for 2h, and then discharged with \gls{cc} \gls{cv} to the voltage corresponding to that \gls{soc} point.
\par
\subsection{Accelerated ageing campaign}\label{subsec:ageing}
The ageing campaign subjected cells to repeated rests, charges, and discharges to measure their effects on performance and provide quantitative results for analysing degradation mechanisms and modes.\par
The ageing campaign comprised 41 cells distributed over four climate chambers (0, 25, and 45°C, and one cycling between 0 and 45°C). The tests span a matrix of temperatures, average \gls{soc}, \gls{dod}, and charge and discharge C-rates, chosen to trigger degradation mechanisms known to occur in Li-ion cells, such as \gls{sei} growth, lithium plating \citep{konz2020detecting}, and pore clogging. \Cref{tab:ageing_matrix} maps each cell to the factor it isolates and to its temperature. The matrix design is described in \Cref{subsec:matrix_design}, while the common monthly \glspl{rpt} and analysis tools used for each cell's experiment are described in \Cref{subsec:RPTs}. The cells were monitored for approximately one year or until cell failure.
\par
The ageing tests were conducted at TNO and AIT\rev{. S}imilar to the characterisation tests, the cells at TNO were cycled in two 12-channel Weiss Technik 81002 climate chambers, each coupled to a Chroma 7100 Cell Cycler (100A per channel, $\pm$5V). Periodically, at \gls{bol}, \gls{mol} and \gls{eol}, \gls{eis} was conducted at 25°C: the cells were kept in the aforementioned climate chambers, the Chroma Cycler channels were disconnected, and the \gls{eis} channels of the Ivium 5000 machine were connected through the climate-chamber routing. At AIT, a similar test setup was implemented using two climatic chambers from Vötsch Industrietechnik GmbH (one VC 4034 and one VT 4060) to thermally condition and stabilise the cells, and two 12-channel Cell Test Units manufactured by AIT (100A per channel, $\pm$5V) for cycling. The \gls{eis} measurements were conducted at AIT using an SP-150 and Booster VMP3 from Bio-logic.
\subsubsection{Ageing matrix design}\label{subsec:matrix_design}
Each cell, or small group of cells, isolates one degradation factor by varying a single condition around a common baseline (\Cref{tab:ageing_matrix}). The campaign is divided into calendar ageing, in which the cell is held at rest, and cyclic ageing, in which it is repeatedly charged and discharged. Both cyclic and calendar ageing depend not only on the test conditions imposed on the cells, but also on the cathode and anode materials and the cell chemistry \citep{chahbaz2021non,laufen2024multi,kassem2012calendar,streck2024self,he2023high,auch2026new,klick2025failure}.
\par
\paragraph{Calendar ageing} Cells were subjected to continuous rest (zero imposed current, at open circuit), interrupted regularly by \gls{rpt} measurements. Because \glspl{rpt} themselves affect ageing \citep{SCHMITT2017183}, a trade-off is needed between measurement frequency and the additional ageing introduced by the tests, which may add cyclic effects to the nominally calendar-ageing results. Measurements were therefore made once per month. \par
At rest, degradation depends mainly on temperature and \gls{soc} \citep{luh2024comprehensive, khan2025calendar}: the higher each is, the faster the cell degrades \citep{Guenther2013ModelbasedIO, SAEED2024102436}. Five calendar-ageing tests were performed on five cells (one per condition). Three cells were held at 100\% \gls{soc} at 0, 25, and 45°C to isolate the temperature effect, whereas two further cells at 10\% and 50\% \gls{soc}, both at 45°C, isolated the effect of \gls{soc} at a fixed temperature. The rest periods were interrupted only for the \glspl{rpt}.
\par
\paragraph{Cyclic ageing} In cyclic ageing, cells are subjected to continuous charging, discharging and rest periods. In this matrix, the cells share a common baseline (25°C, 0.5C\,/\,0.5C, 100\% \gls{dod}, 50\% mean \gls{soc}), from which one factor is varied at a time. This baseline (Cell\_12) was also used for Cell\_56, Cell\_89 and Cell\_93 (started after middle-of-life) to gauge cell-to-cell consistency. Repeating the baseline at 0, 25 and 45°C, and under a temperature that cycles between 0 and 45°C, isolates the temperature effect. The charge rate is then swept while the discharge rate is held at 0.5C: high charge rates at low temperature provoke lithium plating \citep{SMITH2023233118}, whereas charge rates above the manufacturer limit at 25 and 45°C target accelerated \gls{sei} growth. A few cells reverse this asymmetry, applying the higher rate on discharge. At 25°C the depth of discharge is varied between 10 and 100\% at a symmetric 0.5C\,/\,0.5C rate, and at 45°C three cells cycle over a 50\% window centred on 25, 50 and 75\% mean \gls{soc} to isolate the average-\gls{soc} effect. Finally, two cells at 45°C are charged to an elevated limit of 4.45V, one of them with an additional constant-voltage hold (Cell\_22), to provoke cathode and electrolyte degradation as demonstrated with \gls{sem} imaging and \gls{edx} mapping analysis in \Cref{subsec:electrode_composition}.
\par
\paragraph{Dynamic profiles} A further subset is cycled with application-surrogate dynamic load profiles representing two use cases: a stationary-storage profile (stat.) and an automotive profile (auto.). These are applied at fixed temperature (0, 25 and 45°C) and under cycling temperature, and two additional cells alternate between the two profiles over the year. They serve to validate the cell models under realistic operation.
\par
\subsubsection{General experiment design and incorporated analysis tools}\label{subsec:RPTs}
Each cell's unique ageing experiment is wrapped in a common repeating testing structure to allow clean cell-to-cell comparison and degradation analysis. The following elements are applied:
\par
\paragraph{GITT} A \gls{gitt} was performed on each cell at \gls{bol} and \gls{eol} to allow direct comparison of changes in \gls{ocv}. This complements the C/50 discharge and charge performed during each \gls{rpt}. For some cells, \gls{gitt} could not be performed at \gls{eol} because the cell was compromised. Unlike in the characterisation tests, relaxation periods were fixed at 2h.
\par
\begin{figure*}[!t]
    \centering
  \numberedsubfigures
    \subfloat[Characterisation test sequence.]{%
      \includegraphics[]{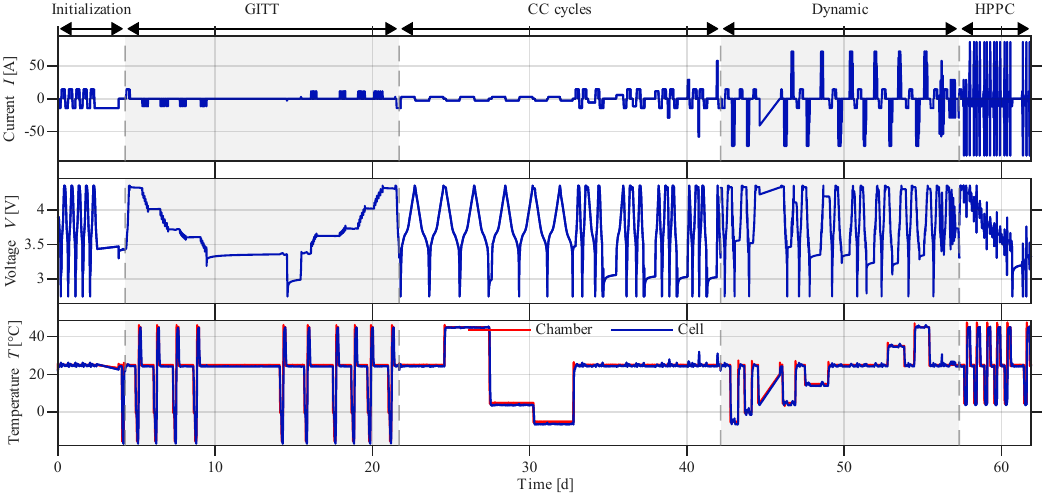}%
        \label{fig:CharacterizationHPPC}%
    }\\[0.75em]
    \subfloat[Monthly reference performance test.]{%
      \includegraphics[]{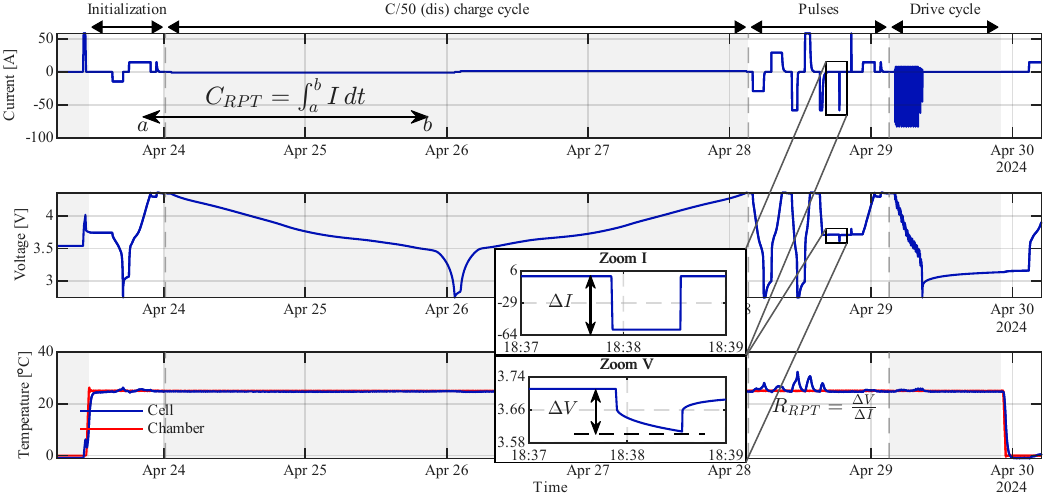}%
        \label{fig:RPTs}%
    }
    \caption{Characterisation and monthly reference performance test data: combined \gls{gitt}, \gls{cc}, dynamic and \gls{hppc} cycles, and the \gls{rpt} current, voltage and temperature profile.}
\end{figure*}

\paragraph{RPT} Each month, a \gls{rpt} was performed to assess changes in performance. Its structure is detailed in \Cref{fig:RPTs}. Key elements include C/50 discharge and charge for dV/dQ or dQ/dV analysis, pulse excitation for resistance tracking, and a drive cycle for tracking changes in impedance across \gls{soc} or for voltage-based model validation.
\par
\paragraph{EIS} At \gls{bol}, \gls{mol}, and \gls{eol}, \gls{eis} sweeps were performed at 90\%, 50\%, and 10\% \gls{soc}. The regime was the same as that used for characterisation, except that these sweeps were performed only at 25°C. In groups A\_0 and A\_25, \gls{eis} was performed on all cells; in A\_45 and A\_var, it was performed on selected cells because of equipment limitations. For some cells, \gls{eis} could not be performed at \gls{mol} and/or \gls{eol} because of early cell failure.
\par
\paragraph{Post-discharge relaxation} After each discharge in the cyclic ageing tests, a 15-minute rest is applied before charging resumes. This allows concentration gradients built up during discharge to relax, largely in the electrolyte and partly in the solid particles, so that each charge begins from a more homogeneous, reproducible state~\citep{rashid2015relaxation}. The rest follows discharge because relaxation at this point affects the following cycle more than relaxation after charge, while the low state of charge, at which the negative electrode is delithiated, keeps additional calendar ageing negligible~\citep{rashid2015relaxation,schreiber2025relaxation}. This ensures that each post-charge stripping measurement starts from a well-relaxed initial state.
\par
\paragraph{Post-charge Li-stripping detection}
A single \gls{rpt} cannot pinpoint when a particular ageing mechanism is triggered or which mechanism drives the observed degradation. However, lithium plating can be followed by lithium stripping, either at rest or during discharge \citep{konz2020detecting}. In dV/dt or dV/dQ analysis, this stripping appears as a characteristic bump and is commonly used as evidence that plating occurred in the preceding step. To exploit this signature, a short discharge (C/5 for 15
minutes) is appended after every charge of the cyclic ageing
cells. The stationary-storage profile is the only exception because
it has no uniform charging section to which the discharge
can be appended. This discharge is also omitted during the BoL
and EoL \gls{gitt} tests and during the RPTs.

\par
\section{Results \& Discussion}\label{sec:results}
This section presents the testing data and discusses example results for the CALB 58Ah graphite NMC 532 cell used in the NextBMS project. 
\subsection{Cell Teardown Findings}
In total, two cells were opened, namely a pristine cell (Cell\_96) and a highly degraded cell (A\_45\_Cell\_22).
\begin{figure*}[!h]
    \centering
    \numberedsubfigures
    \subfloat[Half-cell \gls{gitt} and slow C/50 cycle data.]{%
        \begin{minipage}[t]{0.48\textwidth}
            \includegraphics[]{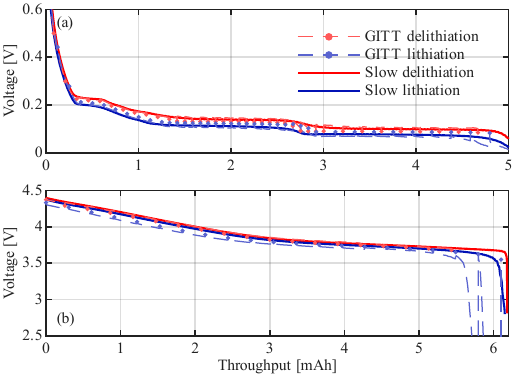}%
        \end{minipage}%
        \label{fig:OCP_HalfCell}%
    }
    \hfill
    \subfloat[Full-cell \gls{ocv} and temperature dependence.]{%
        \begin{minipage}[t]{0.48\textwidth}
            \includegraphics[]{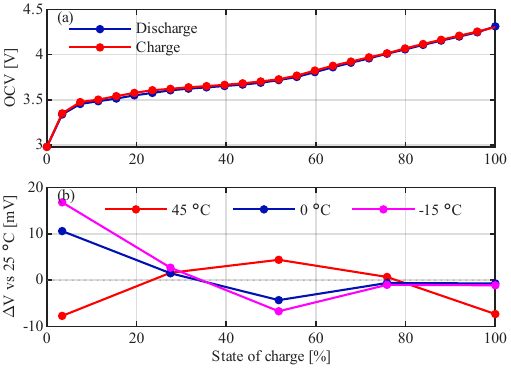}%
        \end{minipage}%
        \label{fig:OCVs_from_GITT}%
    }
    \caption{Half-cell and full-cell open-circuit-voltage data. The half-cell panel shows graphite-anode and NMC 532-cathode \gls{gitt} data with corresponding slow C/50 cycles; the full-cell panel shows charge and discharge \glspl{ocv} and their temperature dependence.}
\end{figure*}
The cell teardown shows that each cell contains two separate, presumably identical jelly rolls (see \Cref{fig:cell_teardown_stages}(b)), each with multiple tabs. The jelly rolls are welded in parallel and connected to the outer cell terminals to optimise the cell's internal resistance, as seen in \Cref{fig:cell_teardown_stages}(d).
Each jelly roll is formed by wound double-coated electrodes (not z-folded).
The total length of a single cathode roll is 5.80m, its width is 8.6cm, and the average thickness of the double-coated electrode with current collector is 115$\mu$m. The cathode current collector is 11.7$\mu$m thick, and the single-sided cathode coating is approximately 52$\mu$m thick.
The single-sided loading of the cathode is 15.97mg/cm$^2$ $\pm$ 0.36mg/cm$^2$.
The total length of a single anode roll is 6.11m, its width is the same as that of the cathode (8.6cm), and the average thickness of the double-coated electrode with current collector is 140$\mu$m. The anode current collector is 6.1$\mu$m thick, and the single-sided anode coating is approximately 67$\mu$m thick.
The single-sided loading of the anode is 9.89mg/cm$^2$ $\pm$ 0.16mg/cm$^2$.
Finally, the thickness of the separator is 15$\mu$m.
\subsubsection{Open Circuit Potentials}\label{subsubsec:ocp}
\par
The half-cell measurements provided three \gls{ocp} measurements for each electrode and direction of (de)lithiation: \gls{gitt} and \gls{cc} at C/50 and C/100.
In general, the \gls{gitt} measurement best reduced overpotential and avoided hysteretic effects, falling between the measurements using \gls{cc} at C/50 and C/100. The practical specific capacity of NMC 532 is taken as 172mAhg$^{-1}$, and the active-material mass of the NMC 532 electrode is 0.03553g. The practical specific capacity of graphite is taken as 300mAhg$^{-1}$, and the active-material mass of the graphite electrode is 0.01668g.
\par
The OCP of the NMC 532 was measured between potential limits of 2.8 and 4.4V vs Li/Li+. For the graphite anode, one full cycle was performed between 0.01 and 2V vs Li/Li+, followed by a \gls{gitt} between 0.01 and 1.5V; \rev{\Cref{fig:OCP_HalfCell}(a)} shows the resulting OCP. \rev{Both electrodes follow the same recipe: a full C/20 conditioning cycle, then titration blocks of C/20 to C/50 pulses (20min to 2.5h per pulse) with 5 to 20h relaxation periods, segmented over the specific-capacity window of the electrode. The complete step-by-step protocols are provided with the dataset in \texttt{2\_HalfCell/OCP\_data/README.md}.}

\subsubsection{Electrode Level Analysis}\label{subsec:electrode_composition}
\begin{figure*}[!t]
    \centering
    \numberedsubfigures
    \subfloat[SEM imaging of anode shown at different magnifications.]{\includegraphics[width=0.48\textwidth,height=0.20\textheight]{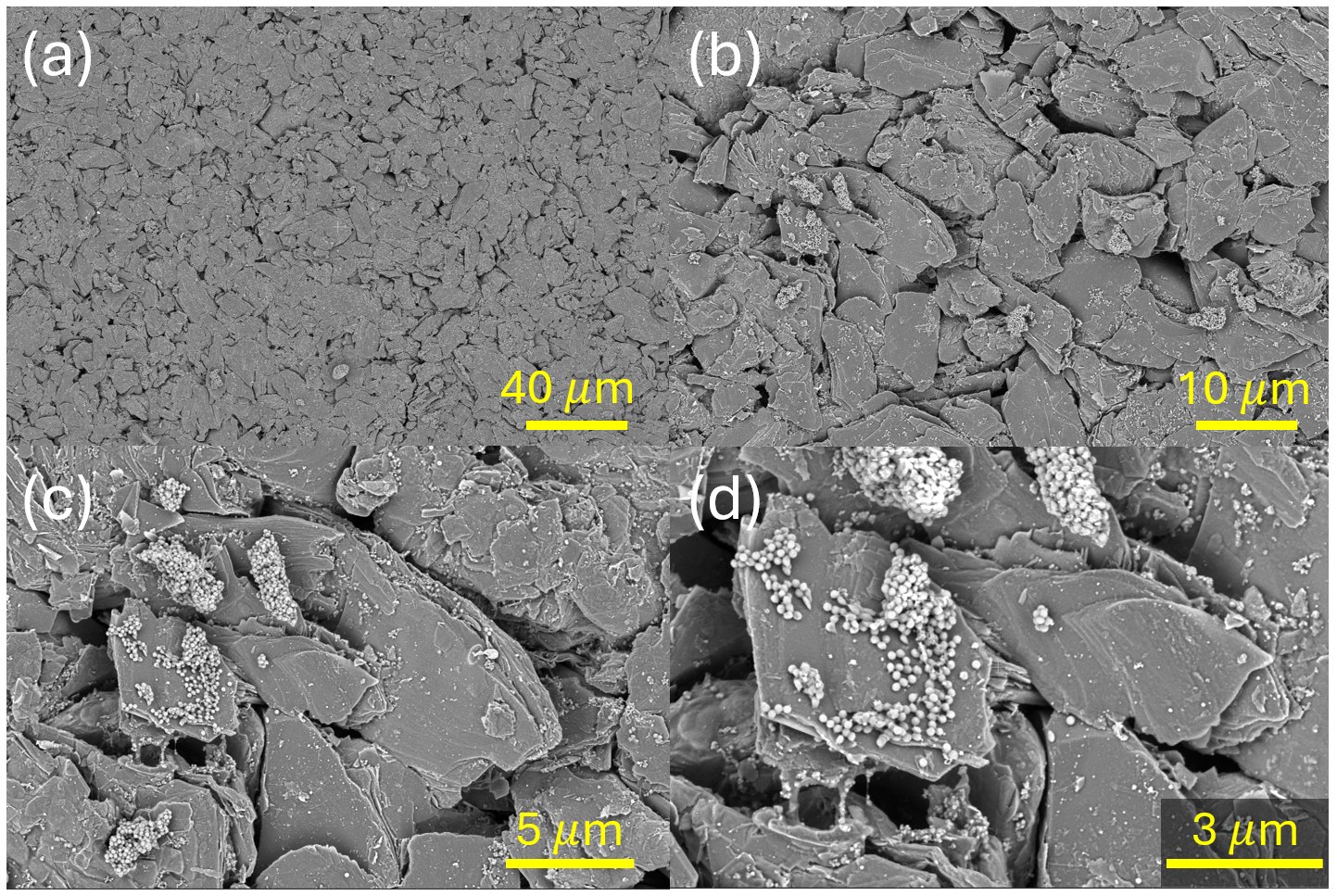}\label{fig:SEM_anode_pristine}}\hfill
    \subfloat[EDX mapping of anode (a) and the corresponding elemental maps: (b) C K$\alpha$1,2, (c) O K$\alpha$1, (d) F K$\alpha$1,2, (e) P K$\alpha$1, (f) S K$\alpha$1.]{\includegraphics[width=0.48\textwidth,height=0.20\textheight]{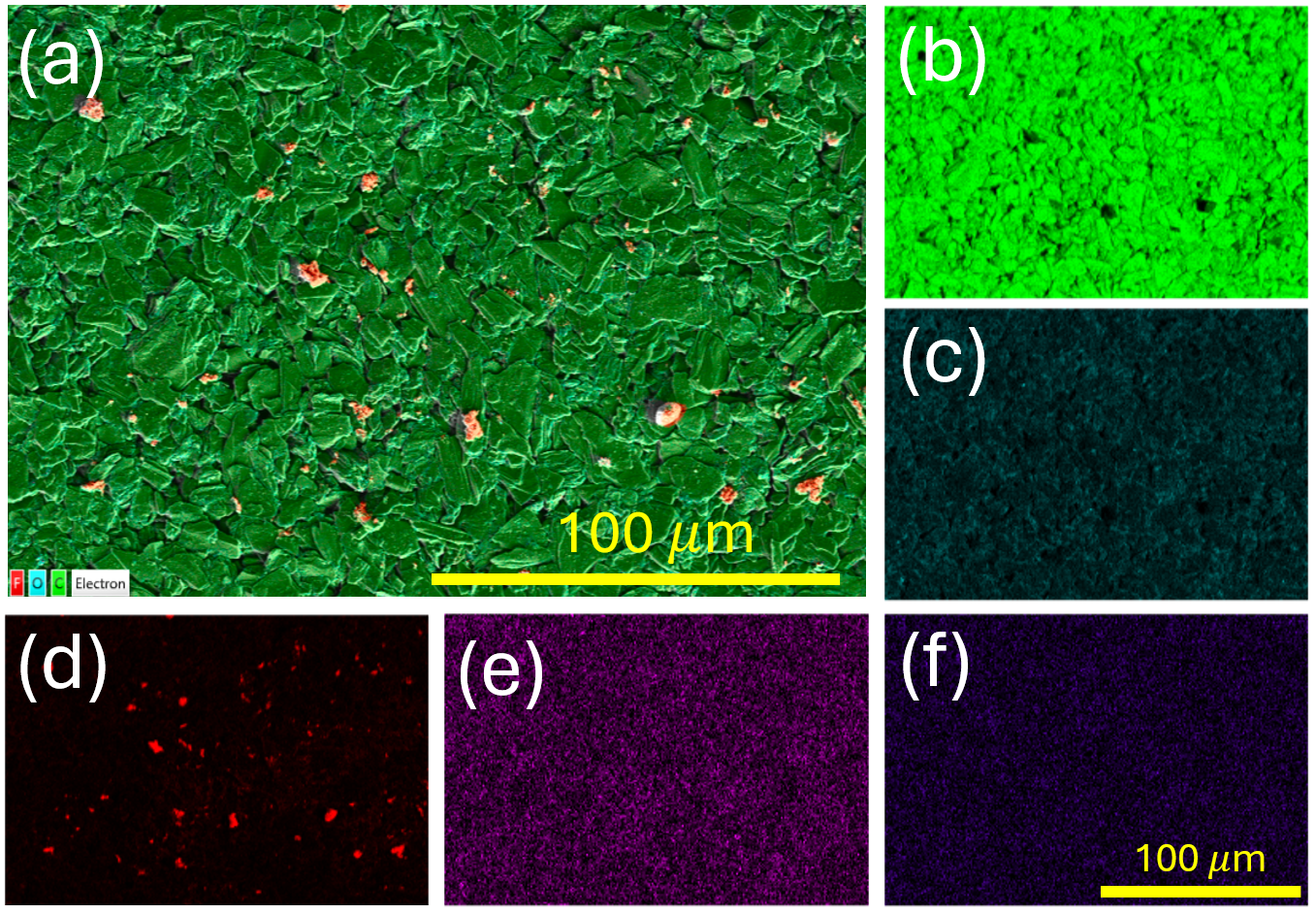}\label{fig:SEM_EDX_pristine_anode}}\\[0.5em]
    \subfloat[SEM imaging of deposits on the surface of the cathode (a-d).]{\includegraphics[width=0.48\textwidth,height=0.20\textheight]{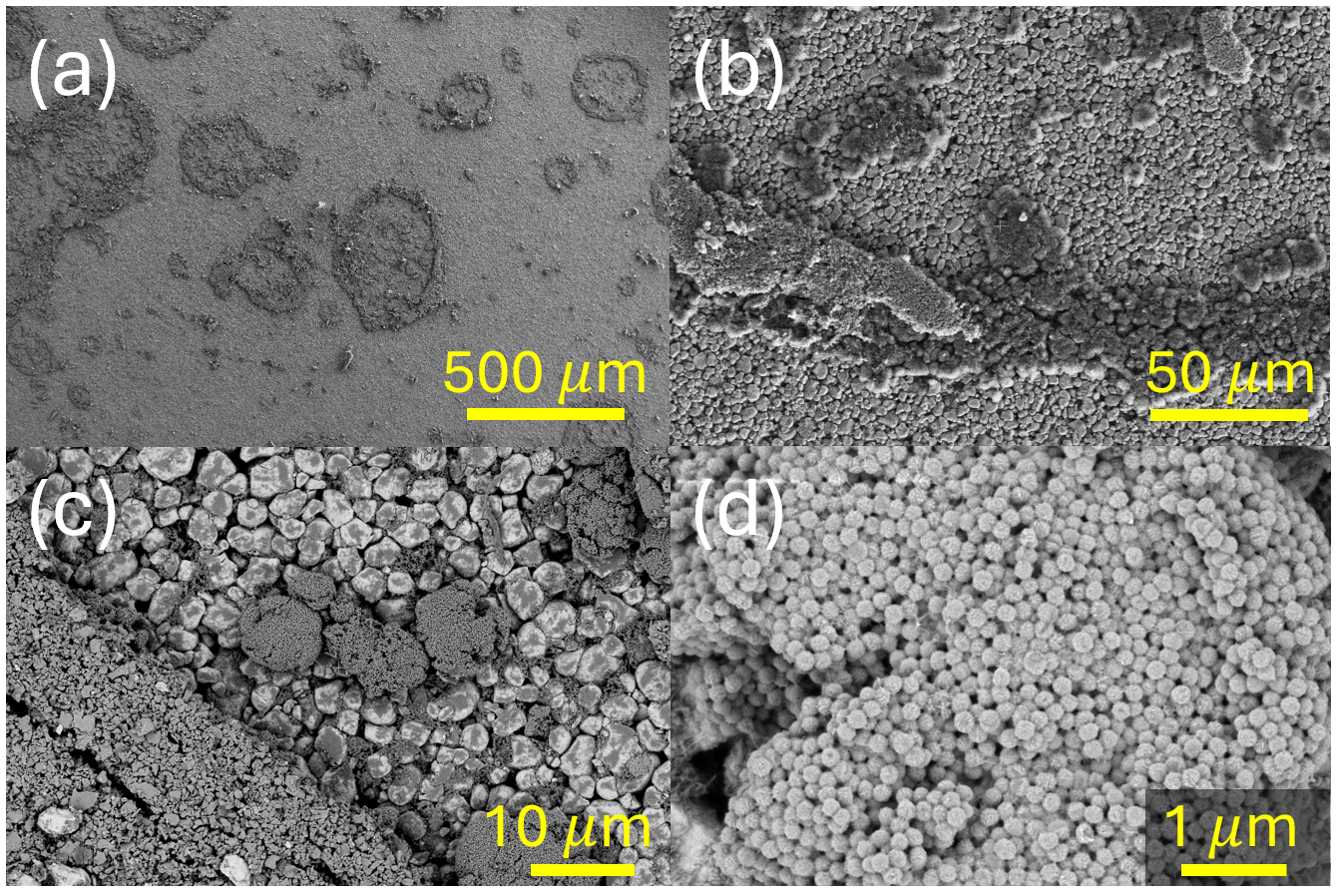}\label{fig:SEM_pristine_cathode_deposits}}\hfill
    \subfloat[EDX mapping of deposits on the surface of the cathode (a) and the corresponding elemental maps: (b) C K$\alpha$1,2, (c) O K$\alpha$1, (d) F K$\alpha$1,2, (e) Al K$\alpha$1, (f) Ni L$\alpha$1,2, (g) Mn L$\alpha$1,2, (h) Co L$\alpha$1,2, and (i) P K$\alpha$1.]{\includegraphics[width=0.48\textwidth,height=0.20\textheight]{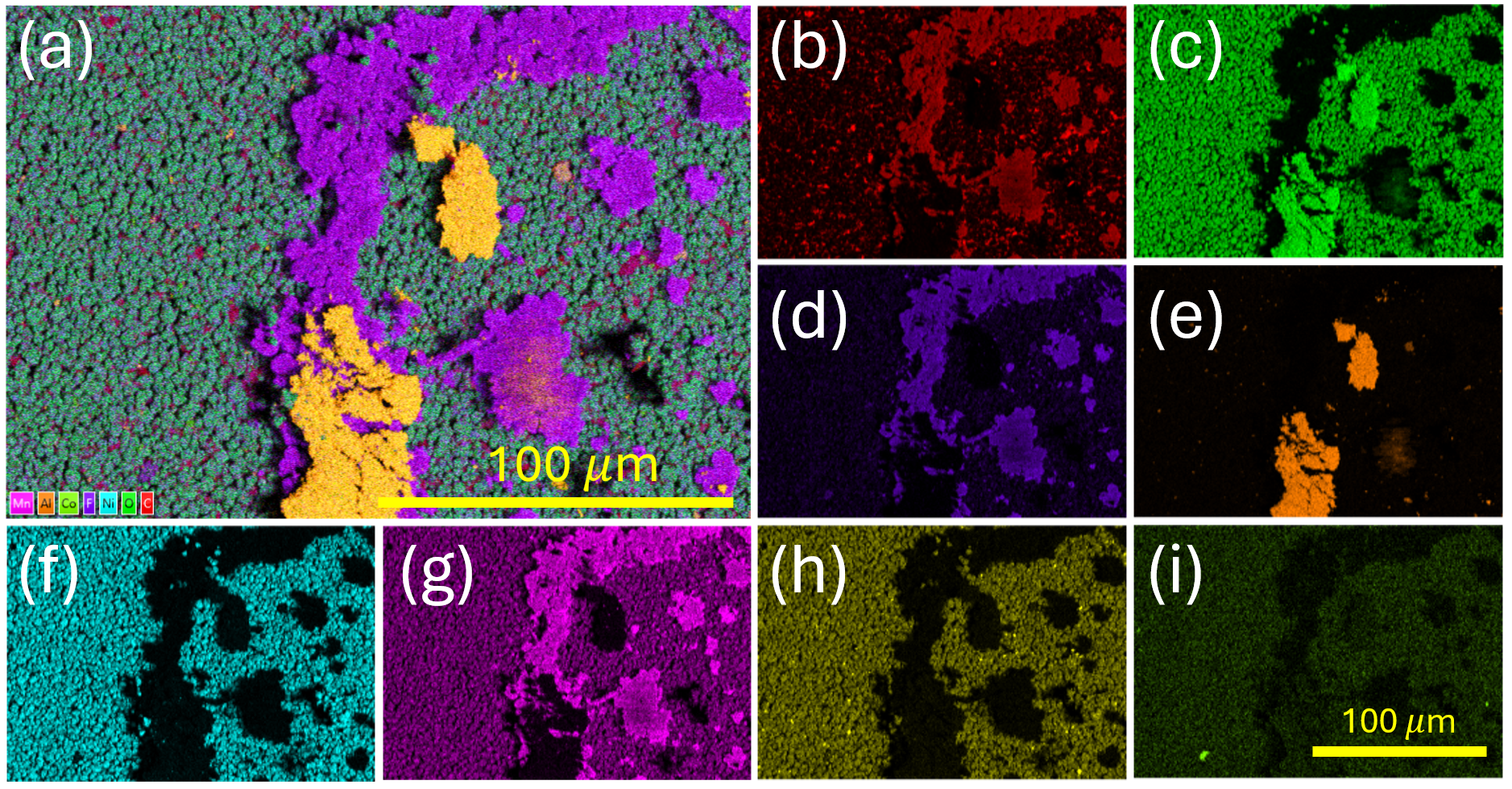}\label{fig:SEM_EDX_pristine_cathode_deposits}}\\[0.5em]
    \subfloat[SEM imaging of cathode active material particles (a-d).]{\includegraphics[width=0.48\textwidth,height=0.20\textheight]{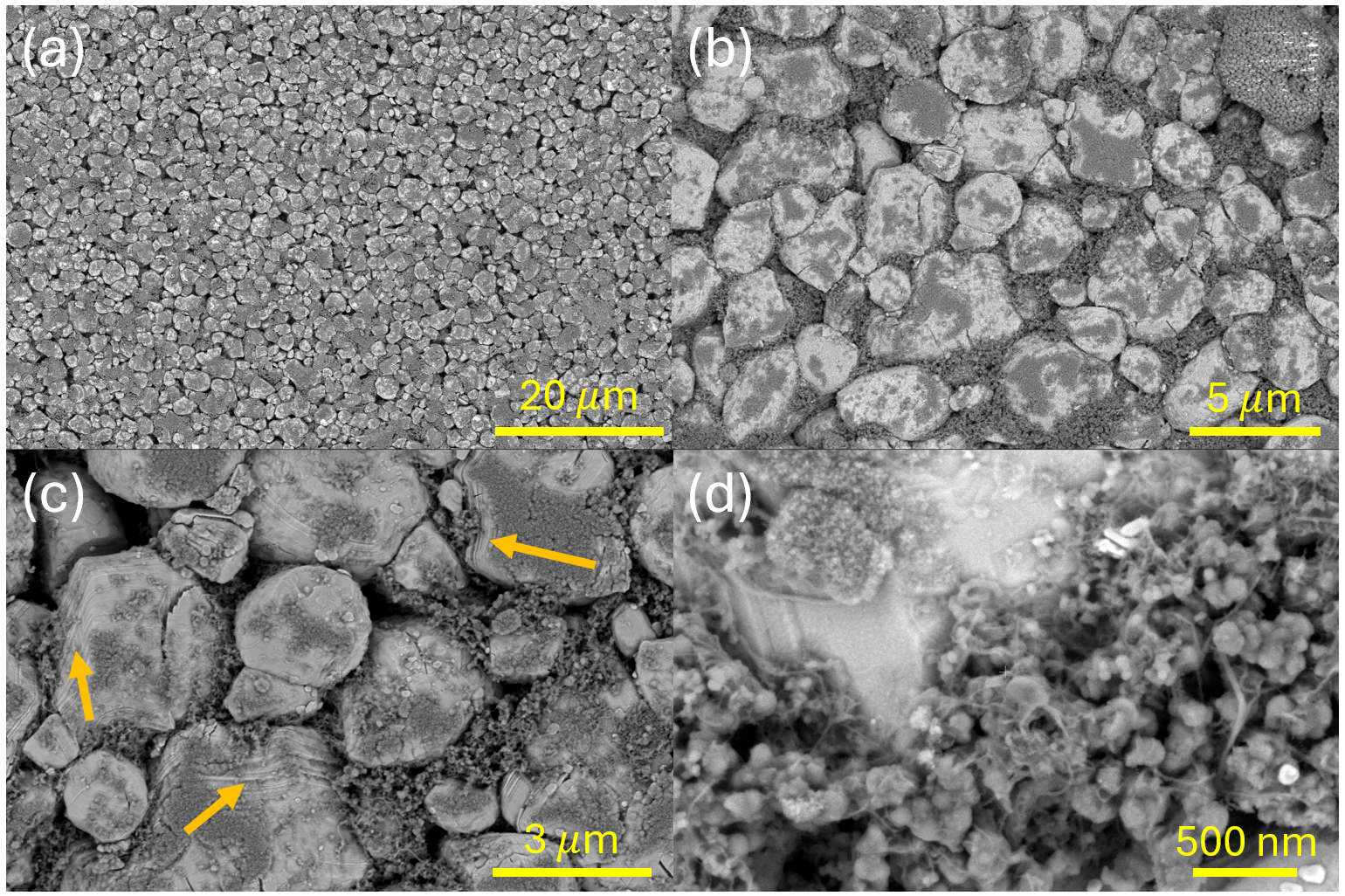}\label{fig:SEM_pristine_cathode}}\hfill
    \subfloat[EDX mapping of cathode active material particles (a) and the corresponding elemental maps: (b) C K$\alpha$1,2, (c) O K$\alpha$1, (d) F K$\alpha$1,2, (e) Al K$\alpha$1, (f) Ni L$\alpha$1,2, (g) Mn L$\alpha$1,2, (h) Co L$\alpha$1,2, and (i) P K$\alpha$1.]{\includegraphics[width=0.48\textwidth,height=0.20\textheight]{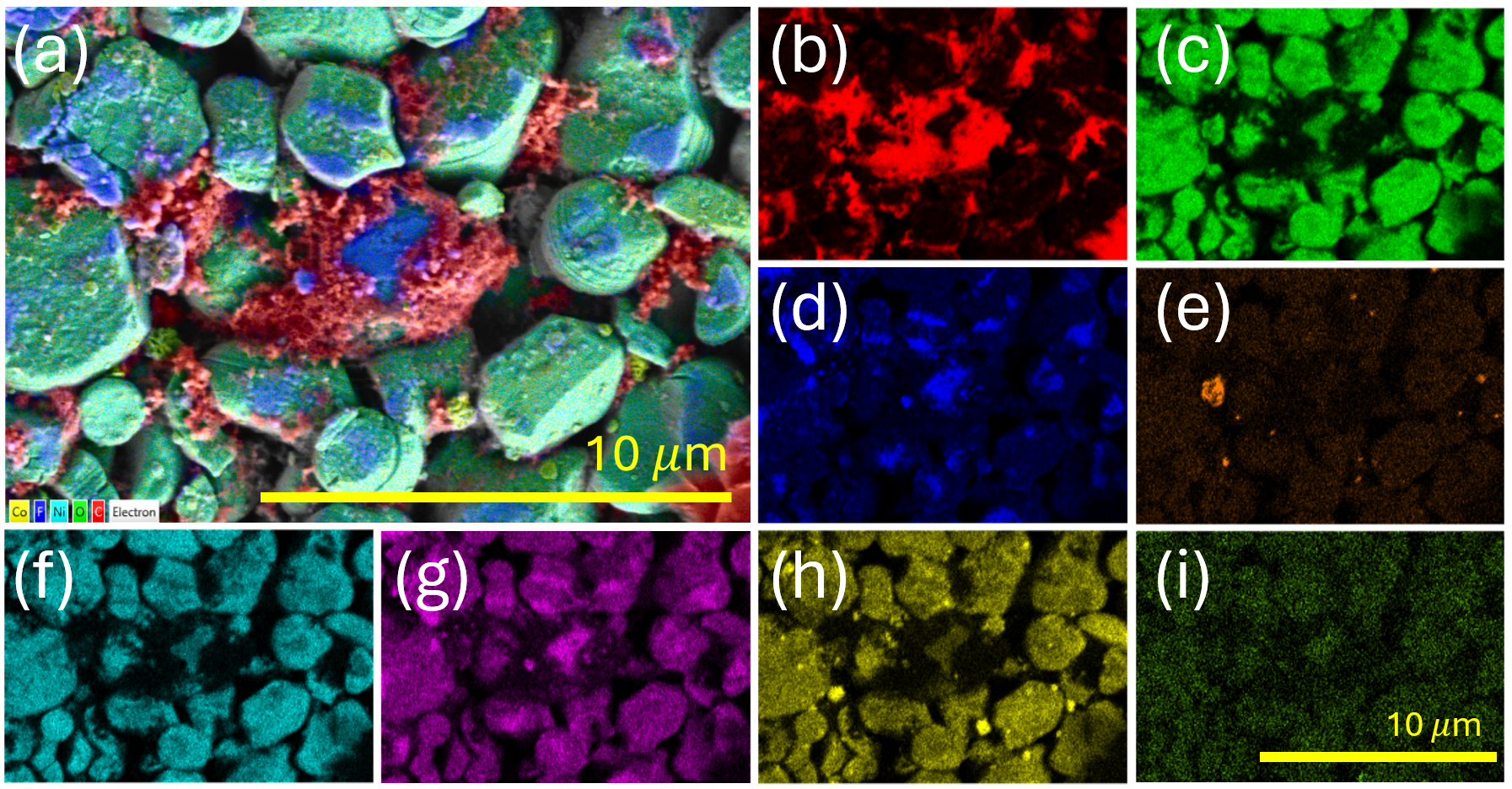}\label{fig:SEM_EDX_pristine_cathode}}
    \caption{Pristine-electrode SEM and EDX montages, including all original anode, cathode-deposit and cathode-active-material images.}
\end{figure*}

\paragraph{Pristine electrode analysis} \gls{sem} imaging revealed the morphology and microstructure of the electrodes.
At low magnification, the surface of the pristine anode extracted from Cell\_96 exhibits a homogeneous and relatively featureless topography (\Cref{fig:SEM_anode_pristine}(a)). Higher-magnification micrographs reveal a predominantly flake-like, non-spherical particle morphology (\Cref{fig:SEM_anode_pristine}(b-d)), with localized clusters of spherical particles adhering to the graphite flake surfaces.
\par
\gls{edx} mapping confirms that the elemental composition of the pristine anode is predominantly carbon, with a minor contribution from oxygen (\Cref{fig:SEM_EDX_pristine_anode}(a)).
Based on the combined morphological characteristics and elemental profile, the active material was identified as synthetic or natural flake graphite.
Additionally, scattered, non-conductive aggregates enriched in fluorine were observed across the graphite surface, suggesting the utilization of PVdF as the polymeric binder.
Trace quantities of sulfur and phosphorus were also detected, likely originating from residual electrolyte salts or functional additives.
Initial low-magnification observations of the pristine cathode surface reveal several circular features reminiscent of dried droplet deposits (\Cref{fig:SEM_pristine_cathode_deposits}(a)).
Higher-magnification micrographs indicate that these deposits have varying surface textures, suggesting multiple distinct secondary phases or surface residues (\Cref{fig:SEM_pristine_cathode_deposits}(b)).
At elevated magnifications, two discrete morphologies within the deposits can be clearly distinguished (\Cref{fig:SEM_pristine_cathode_deposits}(c)): an irregular phase characterized by blocky or rock-like geometries, and a secondary phase composed of large clusters of spherical particles.
As shown in \Cref{fig:SEM_pristine_cathode_deposits}(d), these primary spherical particles have sub-micrometer dimensions and likely correspond to the polymeric binder network.
\par
\begin{figure*}[!h]
    \centering
    \numberedsubfigures
    \subfloat[SEM imaging of the degraded anode (a-d).]{\includegraphics[width=0.48\textwidth]{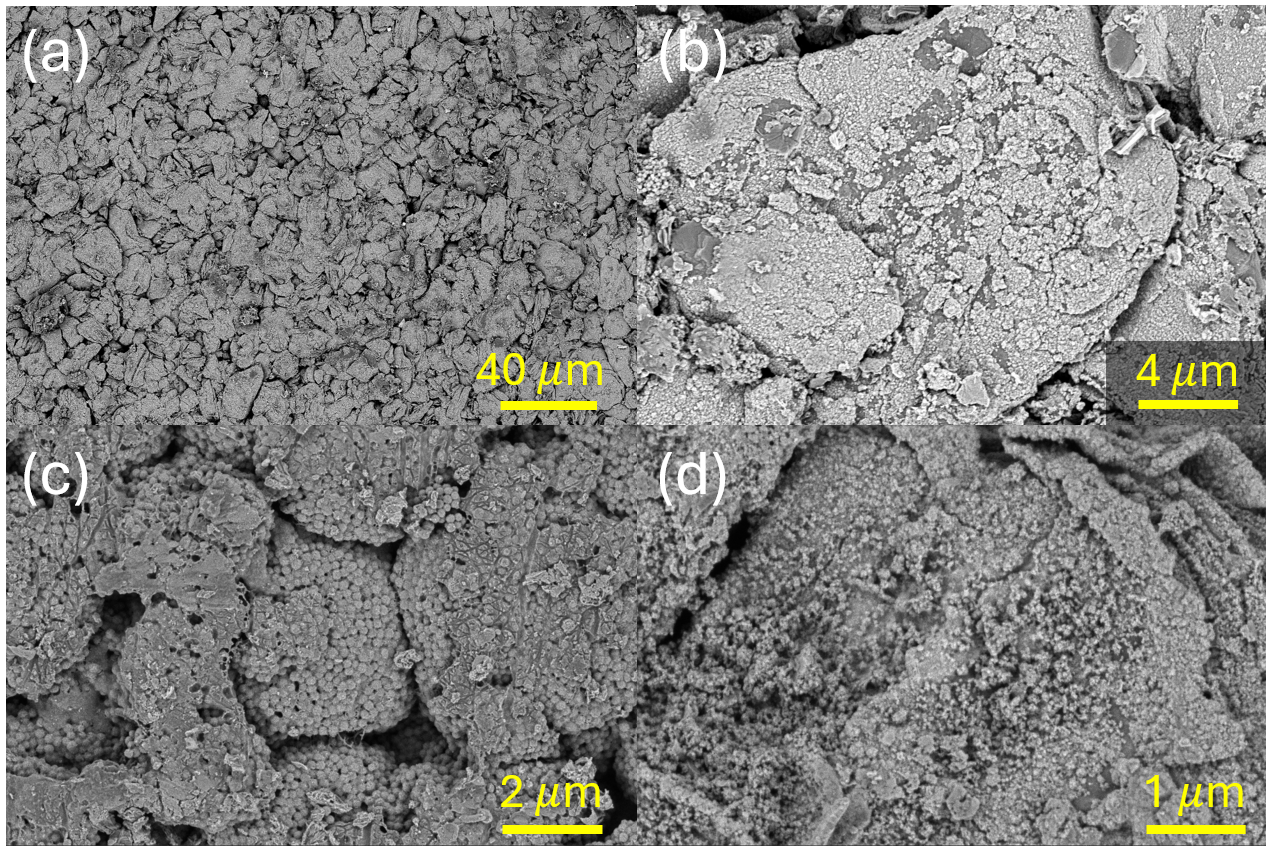}\label{fig:SEM_degraded_anode}}\hfill
    \subfloat[SEM imaging of the degraded cathode (a-d).]{\includegraphics[width=0.48\textwidth]{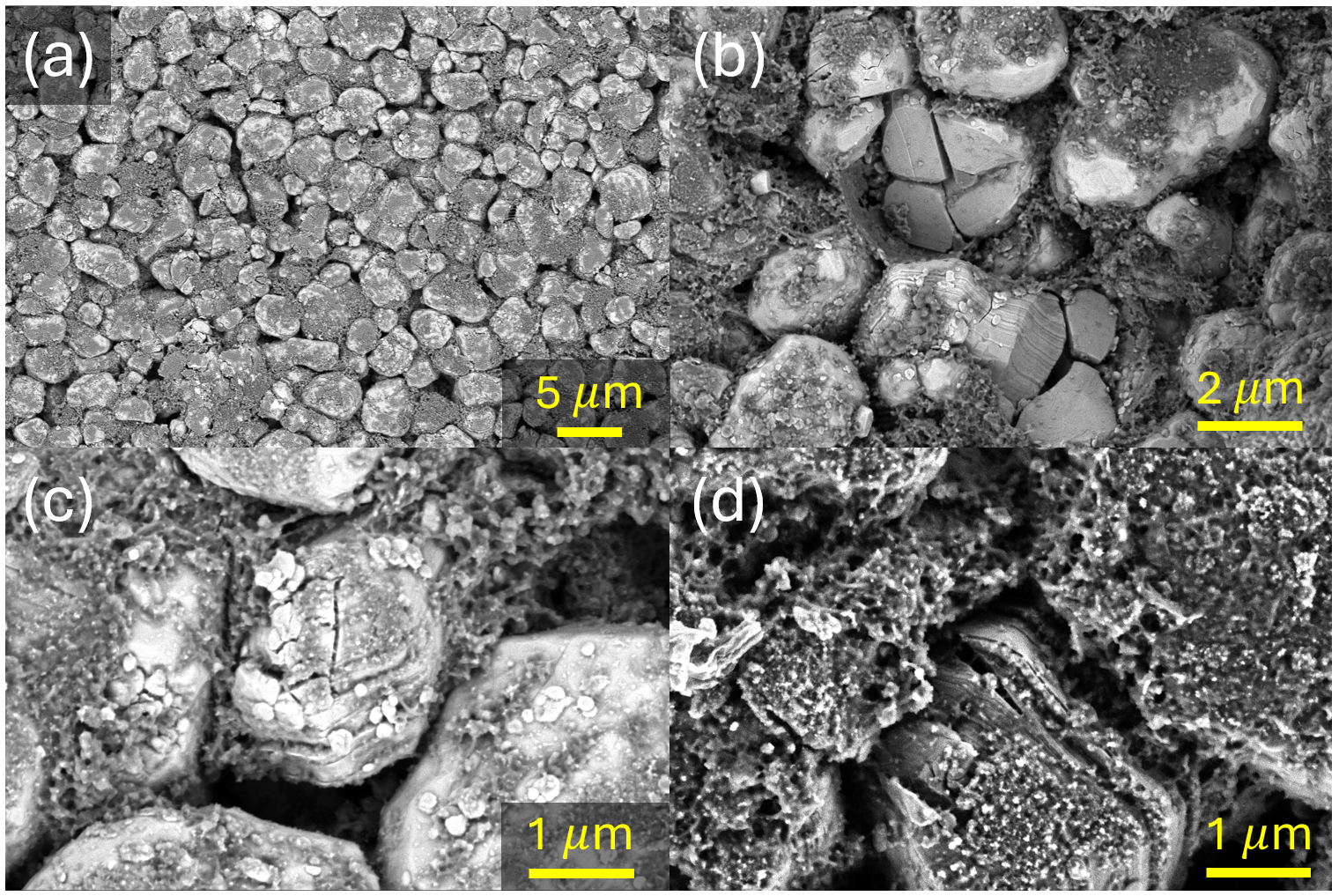}\label{fig:SEM_degraded_cathode}}\\[0.5em]
    \subfloat[EDX mapping of deposits on the surface of the degraded anode (a) and the corresponding elemental maps: (b) C K$\alpha$1,2, (c) O K$\alpha$1, (d) F K$\alpha$1,2, (e) P K$\alpha$1.]{\includegraphics[width=0.48\textwidth,height=0.20\textheight]{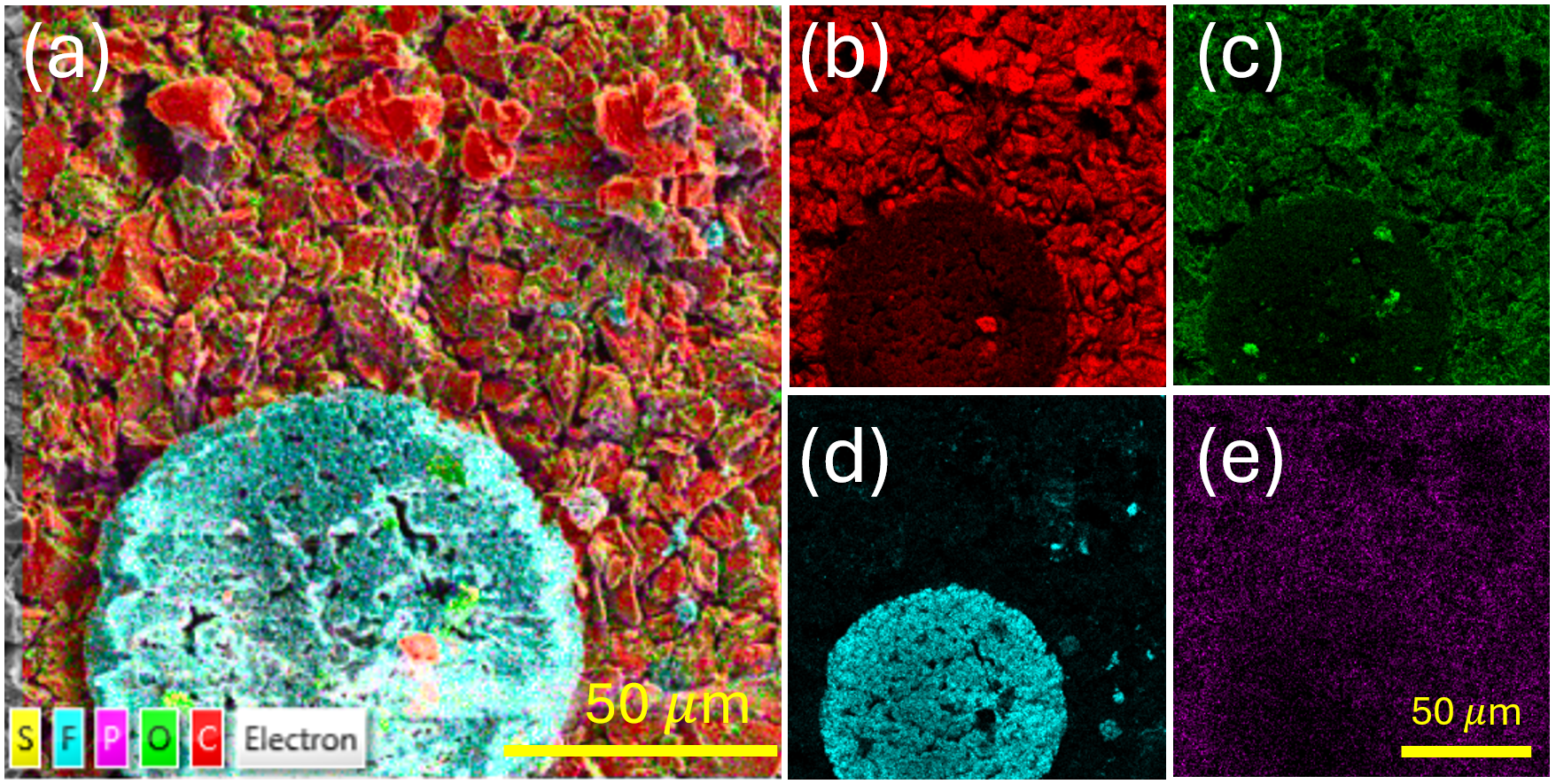}\label{fig:SEM_EDX_degraded_anode_deposits}}\hfill
    \subfloat[EDX mapping of the degraded anode (a) and the corresponding elemental maps: (b) C K$\alpha$1,2, (c) O K$\alpha$1, (d) F K$\alpha$1,2, (e) P K$\alpha$1.]{\includegraphics[width=0.48\textwidth,height=0.20\textheight]{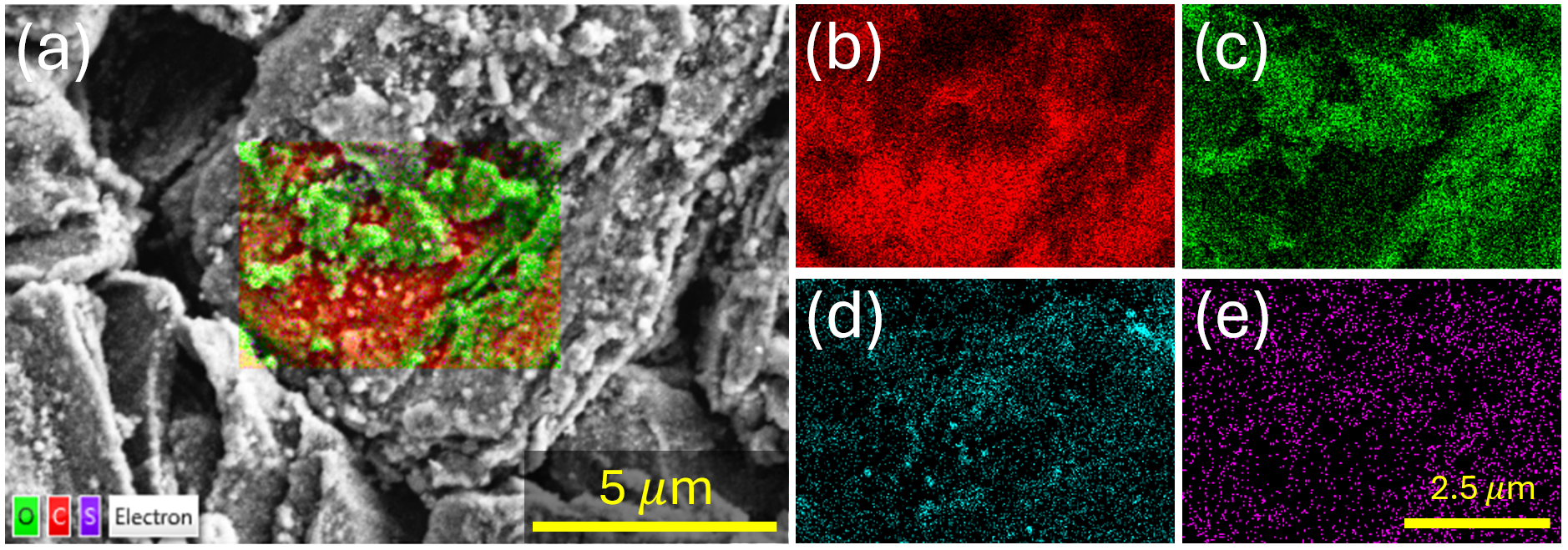}\label{fig:SEM_EDX_degraded_anode}}
    \caption{Degraded-electrode SEM and EDX montages, including the anode, anode surface deposits and cathode images.}
    \label{fig:SEM_EDX_degraded_anode_combined}
\end{figure*}
Localized \gls{edx} spectra and elemental mapping across the deposits reveal the origin and composition of the observed surface structures (\Cref{fig:SEM_EDX_pristine_cathode_deposits}). The analysis revealed dominant signals for aluminum and oxygen, with the corresponding atomic ratios strongly indicating an Al$_2$O$_3$ layer.
This surface alumina is most likely associated with the delamination and subsequent transfer of the Al$_2$O$_3$ coating from the ceramic-coated separator.
Conversely, the underlying layer lacks a well-defined morphology, appearing instead as a diffuse shadow across the cathode active material (CAM) surface.
The \gls{edx} spectrum of this underlying phase (\Cref{fig:SEM_EDX_pristine_cathode_deposits}(d)) exhibits a composition consistent with PVdF, forming a thin, conformal layer that partially encapsulates the CAM particles.
Outside of these localized deposits, the pristine regions composed of CAM particles predominantly consist of Ni, Mn, and Co, alongside carbon species from conductive additives and phosphorus species originating from residual electrolyte decomposition.
Notably, the F K$\alpha$1,2 (\Cref{fig:SEM_EDX_pristine_cathode_deposits}(d)) and Mn L$\alpha$1,2 (\Cref{fig:SEM_EDX_pristine_cathode_deposits}(e)) emission lines overlap significantly due to the proximity of their characteristic X-ray energy levels.
However, given that the fluorine signal is heavily localized at the deposit site while the manganese signal is homogeneously distributed across both the deposit and the surrounding matrix, it can be inferred that the deposit consists primarily of the PVdF binder.
\par
Low-magnification imaging of the pristine cathode extracted from Cell\_96 reveals a homogeneous distribution of NMC particles across the electrode area (\Cref{fig:SEM_pristine_cathode}(a)).
Closer inspection indicates that the active material does not form secondary agglomerates; instead, it consists of discrete primary particles ranging from 1 to 5 µm in size (\Cref{fig:SEM_pristine_cathode}(b)).
At higher magnification, the characteristic layered facets of these primary particles become distinctly visible (\Cref{fig:SEM_pristine_cathode}(c), orange arrows).
Under high-magnification conditions (40 kx, \Cref{fig:SEM_pristine_cathode}(d)), auxiliary electrode components are resolved, manifesting as ultrafine spherical particles embedded within an interconnected fibrous network.
These morphological features are attributed to the integration of carbon nanotubes (CNTs) within the carbon black conductive matrix.
Finally, elemental composition was evaluated using both localized \gls{edx} point analysis and area scans on the CAM particles.
Both configurations confirmed Ni, Mn, and Co as the predominant transition metal constituents.
\par
Quantification of the spectra yielded an empirical Ni:Mn:Co atomic ratio of 55:30:15.
Consequently, the cathode active material was precisely identified as Li(Ni$_{0.55}$Mn$_{0.30}$Co$_{0.15}$)O$_2$ (NMC532).
A comprehensive assessment of the composite cathode architecture is provided by high-magnification \gls{edx} elemental mapping (\Cref{fig:SEM_EDX_pristine_cathode}(a)).
The predominant signals correspond to Ni, Mn, and Co (\Cref{fig:SEM_EDX_pristine_cathode}(f-h)), as well as O (\Cref{fig:SEM_EDX_pristine_cathode}(c)), reflecting the characteristic elemental signature of the NMC active material.
Notably, localized variations in transition metal distribution are observed, with certain sub-micrometer particles exhibiting distinct Co-enrichment.
Within the interstitial voids between the primary CAM particles, a pronounced signal for carbon is detected (\Cref{fig:SEM_EDX_pristine_cathode}(b)), which is attributed to the carbon black conductive matrix and, more specifically, the interconnected carbon nanotube network identified in \Cref{fig:SEM_pristine_cathode}(d).
Furthermore, the presence of fluorine (\Cref{fig:SEM_EDX_pristine_cathode}(d)) and phosphorus (\Cref{fig:SEM_EDX_pristine_cathode}(i)) species across the CAM surfaces indicates the accumulation of residual electrolyte decomposition products.
Interestingly, the fluorine contrast is significantly enhanced at localized surface deposits.
Because phosphorus is entirely absent within these specific regions, this localized fluorine signal cannot be attributed to electrolyte salt residues; instead, it underscores the presence of the PVdF binder phase.
Finally, isolated Al-rich domains are detected embedded within the electrode matrix (\Cref{fig:SEM_EDX_pristine_cathode}(e)).
This further supports the hypothesis that the ceramic Al$_2$O$_3$ coating from the separator underwent mechanical delamination, resulting in the transfer of fragments onto the cathode surface.
\par
\paragraph{Aged electrode analysis} Morphological evaluation of the cycled anode extracted from Cell\_22 reveals a significantly altered surface topography, exhibiting a flattened or smeared appearance compared to the pristine electrode (\Cref{fig:SEM_degraded_anode}(a)).
At higher magnifications, the formation of a dense, crust-like passivation layer, consistent with an extensive \gls{sei}, is observed entirely covering the graphite flakes (\Cref{fig:SEM_degraded_anode}(b)).
This heterogeneous surface layer comprises multi-sized particulates and exhibits prominent micro-fissures, exposing isolated domains of the underlying pristine graphite surface.
Additionally, regions containing dense aggregates of spherical particles are visible (\Cref{fig:SEM_degraded_anode}(c)), which are similarly encapsulated by irregular degradation products.
High-magnification microscopy (\Cref{fig:SEM_degraded_anode}(d)) further elucidates the fine structure of these surface deposits, revealing that the macro-scale crust observed in \Cref{fig:SEM_degraded_anode}(b) is actually an agglomeration of primary sub-micrometer particulates densely covering the graphite flake facets.
\par
\gls{edx} mapping on the degraded Cell\_22 reveals the chemical composition of the degradation products covering the cycled anode surface (\Cref{fig:SEM_EDX_degraded_anode}(a)), with a pronounced enrichment in oxygen, fluorine, and phosphorus (\Cref{fig:SEM_EDX_degraded_anode}(c-e)), confirming that the surface deposits originate from the decomposition of the electrolyte salts and organic solvents.
Due to the dense and continuous coverage of this thick passivation layer across the electrode topography, potential structural degradation of the active material, such as graphite exfoliation, could not be conclusively determined via top-down SEM observation.
Apart from the continuous passivation layers covering the graphite surface, distinct circular deposits were observed exclusively on the cycled electrode topography (\Cref{fig:SEM_EDX_degraded_anode_deposits}(a)).
\gls{edx} mapping identified these localized features as agglomerated PVdF binder domains, with fluorine-rich regions providing sharp contrast in the elemental maps (\Cref{fig:SEM_EDX_degraded_anode_deposits}(d)).
The reduction in the carbon signal within these specific zones is a relative effect, reflecting the contrast attenuation against the surrounding carbon-dense graphite matrix rather than an absolute depletion of carbon species.
Furthermore, these macro-features are unrelated to electrolyte decomposition products, as evidenced by the negligible phosphorus signal detected within the domains.
This spatial segregation suggests potential binder migration or phase separation during extended electrochemical cycling.
Microscopic evaluation of the cycled cathode extracted from Cell\_22 reveals surface deposits structurally analogous to those identified on the pristine electrode, albeit with a significantly higher surface coverage across the CAM particles (\Cref{fig:SEM_degraded_cathode}(a)).
\par
High-magnification imaging of the cycled cathode unveils severe mechanical degradation, manifested as pronounced microcracking within numerous CAM particles (\Cref{fig:SEM_degraded_cathode}(b)).
Although isolated fractured particles were present in the pristine state, the affected grains in the cycled sample exhibit pronounced structural fragmentation; this cracking is not observed in every particle.
Furthermore, sub-micrometer fragments, morphologically identical to the primary CAM phase, are resolved adhering to the facets of larger host particles (\Cref{fig:SEM_degraded_cathode}(c)).
This topology strongly indicates that severe particle pulverization occurred during cycling, leading to the detachment and subsequent surface migration of these fractured remnants.
Finally, utilizing an alternative SEM imaging mode reveals that the CAM surfaces are uniformly decorated with ultra-fine, spherical secondary deposits (\Cref{fig:SEM_degraded_cathode}(d)).

\subsection{Cell Characterization Results}\label{subsec:characterization_results}
\begin{figure}[!t]
    \centering
    \numberedsubfigures
    \subfloat[EIS spectra at \gls{bol}.]{%
        \includegraphics[]{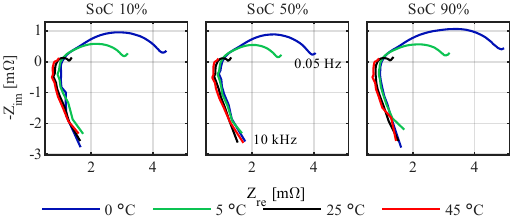}%
        \label{fig:EIS}%
    }\\[0.75em]
    \subfloat[Li-stripping analysis.]{%
        \includegraphics[]{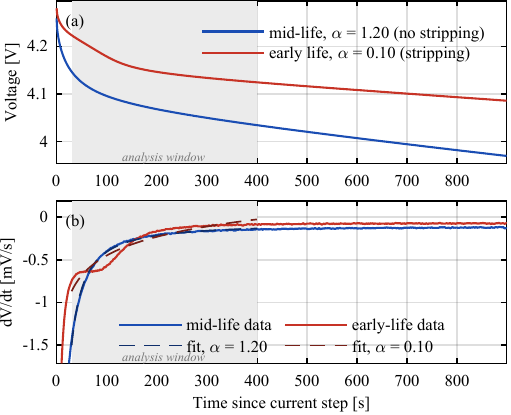}%
        \label{fig:Li-stripping}%
    }
    \caption{Electrochemical impedance and Li-stripping results: EIS spectra at \gls{bol} as functions of \gls{soc} and temperature, and the post-charge C/5 discharge segments and power-law fits used to derive the per-cycle stripping marker.}
\end{figure}

The characterization data provide information useful for benchmarking, cell selection, performance verification, and model development. Examples include detailed discharge and charge \gls{ocv} curves and their temperature dependence from the \gls{gitt} experiment outlined in \Cref{fig:CharacterizationHPPC} and shown in \Cref{fig:OCVs_from_GITT}, as well as \gls{eis} spectra as functions of \gls{soc} and temperature (\Cref{fig:EIS}).
\par
To exploit the post-charge lithium-stripping signature described in \Cref{subsec:RPTs}, we fit a power law \(a + b\,t^{-\alpha}\) to the \(\mathrm{d}V/\mathrm{d}t\) trace over the window \([t_\mathrm{start},t_\mathrm{end}]\), and take the residual
\begin{equation}\label{eq:power-law}
\varepsilon \;=\; \min_{a,\,b,\,\alpha}\;
\sum_{t_i = t_\mathrm{start}}^{t_\mathrm{end}}
\left[\frac{\mathrm{d}V}{\mathrm{d}t}(t_i) - \left(a + b\,t_i^{-\alpha}\right)\right]^{2}.
\end{equation}
\begin{figure*}[!h]
    \centering
    \numberedsubfigures
    \subfloat[Calendar ageing.]{%
        \begin{minipage}[t]{0.48\textwidth}
            \includegraphics[]{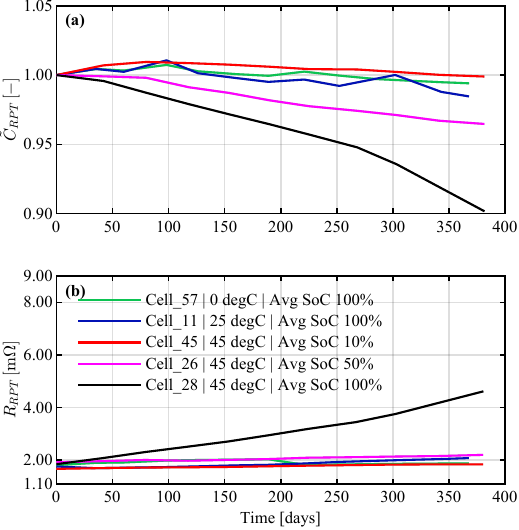}%
        \end{minipage}%
        \label{fig:CalendarAgeing}%
    }
    \hfill
    \subfloat[Cyclic ageing.]{%
        \begin{minipage}[t]{0.48\textwidth}
            \includegraphics[]{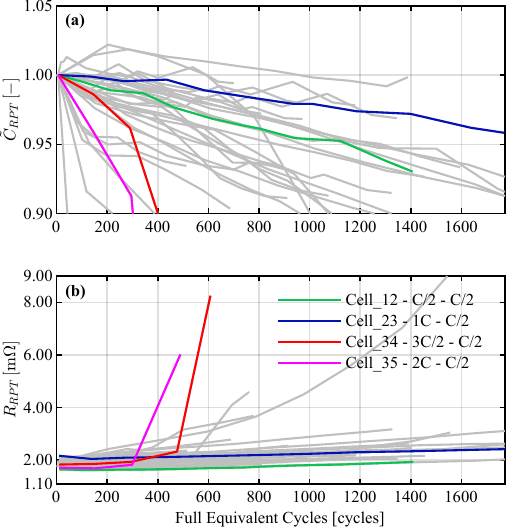}%
        \end{minipage}%
        \label{fig:CyclicAgeing}%
    }
    \caption{Calendar- and cyclic-ageing capacity and resistance trajectories. Calendar trajectories are shown versus time; cyclic trajectories are shown versus full equivalent cycles (FEC).}
    \label{fig:AgeingTrajectories}
\end{figure*}
If no Li stripping occurs, the power law describes the trace well and the residual is low; if stripping occurs, the bump cannot be captured and the residual is high. By fitting the power law to each post-charge discharge and computing the residual, we obtain a time-resolved proxy for identifying which cycle induced plating. A visual demonstration is provided in \Cref{fig:Li-stripping}.

\subsection{Cell Ageing Results}\label{subsec:ageing_degradation}
Rather than present all the data, this section illustrates what readers can obtain from the dataset. For an overview, \Cref{fig:CalendarAgeing} shows the capacity and resistance trajectories for all cells subjected to calendar ageing, and \Cref{fig:CyclicAgeing} shows those for cyclic ageing; both are derived from the \glspl{rpt} described in \Cref{fig:RPTs}. The ageing matrix (\Cref{tab:ageing_matrix}) produced a wide range of trajectories. The severity of ageing followed expectations: high temperature combined with high \gls{soc} produced the most severe calendar ageing, whereas the highest charging C-rates produced severe cyclic ageing.

\begin{figure*}[!h]
    \centering
    \includegraphics[]{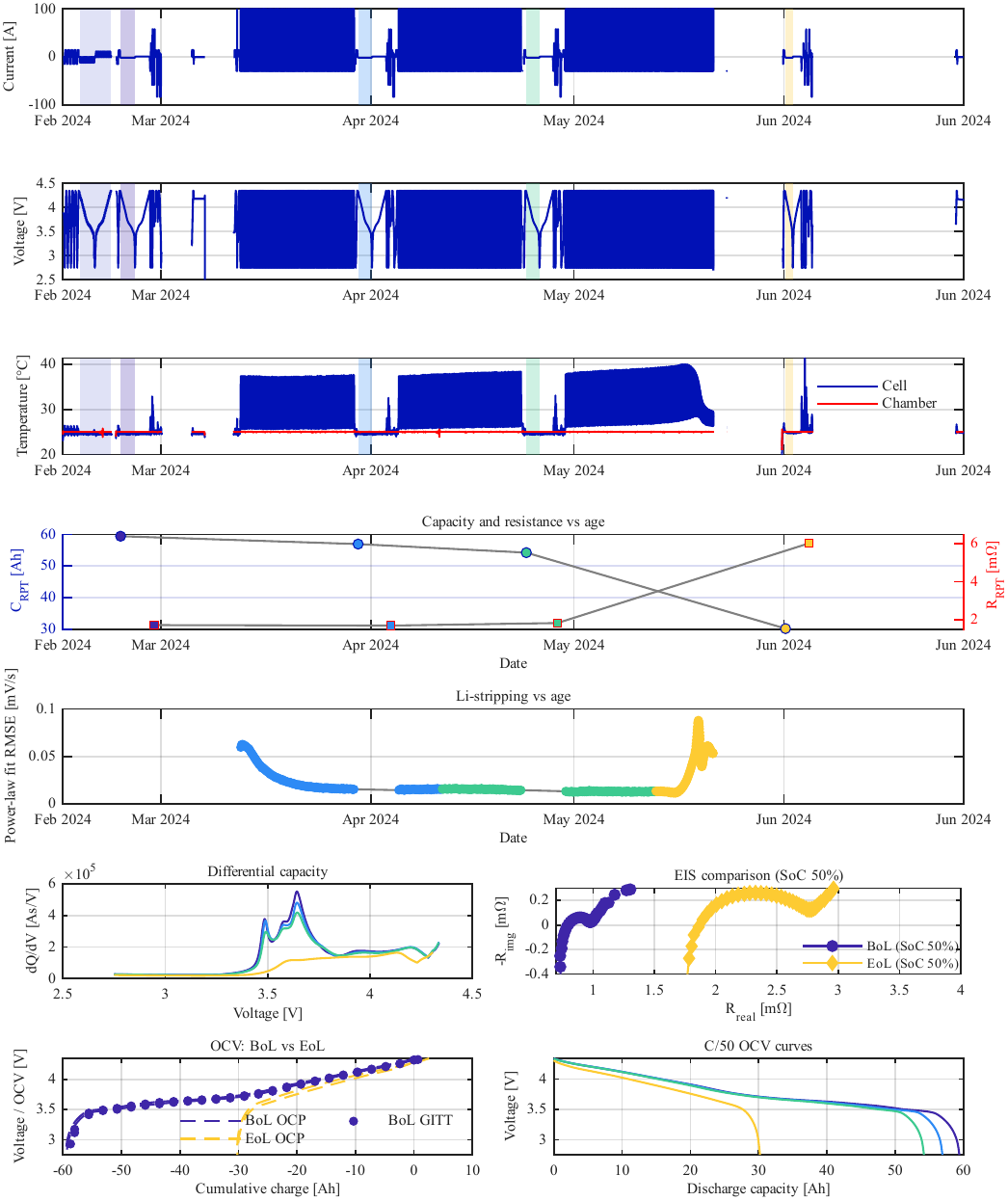}
        \caption{Cell summary for \texttt{Cell\_35}. Coloured bars in the three upper panels mark successive \glspl{rpt}; curves and markers in the lower panels are obtained from these \glspl{rpt}. Matching colours are used to identify the corresponding \gls{rpt} time points. No \gls{eol} \gls{gitt} curve is shown because severe cell degradation left too few measured points.}
     \label{fig:exampleAgeingData}
 \end{figure*}
The per-cell data provide further insights. \Cref{fig:exampleAgeingData} shows an example for Cell\_35, also highlighted in \Cref{fig:CyclicAgeing}. Raw time-series data are shown alongside segments relevant for analysis, obtained from the \glspl{rpt} and other tests, as indicated by the shaded colours. These include the C/50 discharge and charge in each \gls{rpt}, which enable degradation-mode analysis over time, and \gls{eis} data at \gls{bol}, \gls{mol}, and \gls{eol}, where available. Cell\_35 was subjected to a high charging current exceeding datasheet limits. The residual method described in \Cref{subsec:RPTs} detected Li stripping at \gls{bol}; the signal then appeared to self-limit as the test progressed, possibly because of ageing. After a long period without detected stripping, a sudden-death event was accompanied by another peak in the stripping signal. These data can support cycle-to-cycle validation of electrochemical models that include side reactions such as Li plating. In contrast, existing datasets allow side-reaction fits to be evaluated only from accumulated ageing as measured by \gls{rpt} results.

\par
\section{Data Structure}\label{sec:data_structure}
\par
To help readers use and understand the dataset, we describe its repository structure here. The \texttt{DataOverview.md} file at the dataset root \rev{also documents the exact CSV time-column format (an absolute timestamp on the first data row followed by per-row increments in seconds) and per-site formatting differences}.
On Zenodo, the dataset is contained in the folder
\textit{Data from "Characterization, ageing and post-mortem dataset of commercial 58Ah prismatic NMC cells"} and divided into the following top-level subfolders, which follow the structure of \Cref{Sec:Methods}:
\par
\begin{enumerate}[itemsep=0pt, topsep=2pt, parsep=0pt, partopsep=2pt]
  \item \texttt{1\_Teardown/}: cell teardown images, \gls{sem} data and EDX data.
  \item \texttt{2\_HalfCell/}: half-cell \gls{ocp} data for the graphite anode and NMC532 cathode.
  \item \texttt{3\_Characterization/}: full-cell electrical characterisation and \gls{eis} data.
  \item \texttt{4\_Ageing/}: calendar ageing, cyclic ageing and ageing \gls{eis} data.
  \item \texttt{5\_Documentation/}: supporting documentation and the cell datasheet.
\end{enumerate}
\par
Each high-level folder contains data for direct use: raw data in \texttt{.csv} format, SEM data in TIFF format, or pictures in \texttt{.jpg} format. Where available, the data folders also include \texttt{.jpg} figures that visualise the corresponding data; these figures are generated by the plotting scripts shared alongside this paper. 
\par
Calendar and cyclic ageing data are provided in their respective ageing-data folders. Within each ageing folder, there is one subfolder per tested cell, with overview files and processing outputs stored alongside the cell folders. README files within the folders provide further guidance on their contents.
File naming follows the pattern described in \Cref{Sec:Methods}. This convention enables programmatic parsing for large-scale analysis workflows. Examples of the contents and data fields of each subfolder in the Zenodo data are provided below:
\par
\begingroup
\emergencystretch=2em
\raggedright
\begin{itemize}
\item \path{1_Teardown/Teardown_images/Pristine/}\textemdash{} Cell\_96 disassembly images and geometry documentation of a pristine cell, including pictures of the disassembly process and individual cell components.
\item \path{1_Teardown/Teardown_images/Aged/}\textemdash{} Cell\_22 disassembly images of an aged cell, including pictures of the disassembly process and individual cell components.
\item \path{1_Teardown/SEM_data/Pristine/}\textemdash{} scanning-electron-microscopy micrographs of pristine Cell\_96 anode and cathode samples.
\item \path{1_Teardown/SEM_data/Aged/}\textemdash{} scanning-electron-microscopy micrographs of aged Cell\_22 anode and cathode samples.
\item \path{1_Teardown/EDX_data/Pristine/}\textemdash{} energy-dispersive X-ray spectroscopy reports and the pristine-cell composition spreadsheet.
\item \path{1_Teardown/EDX_data/Aged/}\textemdash{} energy-dispersive X-ray spectroscopy reports for the aged-cell samples.
\item \path{2_HalfCell/OCP_data/Anode_Graphite/}\textemdash{} graphite-anode half-cell GITT and slow charge/discharge OCP files, including raw time-series data with Time [s], Current [A] and Voltage [V].
\item \path{2_HalfCell/OCP_data/Cathode_NMC532/}\textemdash{} NMC532-cathode half-cell GITT and slow charge/discharge OCP files, including raw time-series data with Time [s], Current [A] and Voltage [V].
\item \path{3_Characterization/<Cell>/}\textemdash{} per-cell full-cell time-series files for Initialization, GITT, CC, dynamic-cycle and HPPC characterisation, with Time [s], Voltage [V], Current [A], cell temperature [°C] and chamber temperature [°C].
\item \path{3_Characterization/<Cell>/EIS/}\textemdash{} per-cell characterisation EIS data, provided both as recorded time-series data with Time [s], Current [A] and Voltage [V], and as processed spectra with real impedance [$\Omega$], imaginary impedance [$\Omega$] and frequency [Hz] at 10\%, 50\% and \rev{90\%} SoC.
\item \path{4_Ageing/Calendar_ageing_data/}\textemdash{} per-cell calendar-ageing time-series data and associated processing files, with Time [s], Voltage [V], Current [A], cell temperature [°C] and chamber temperature [°C]. The corresponding storage conditions are summarised in the Methods ageing matrix (\Cref{tab:ageing_matrix}).
\item \path{4_Ageing/Cyclic_ageing_data/}\textemdash{} per-cell cyclic-ageing time-series data and associated processing files, with Time [s], Voltage [V], Current [A], cell temperature [°C] and chamber temperature [°C]. The corresponding cycling conditions are summarised in the Methods ageing matrix (\Cref{tab:ageing_matrix}).
\item \path{4_Ageing/EIS_data/}\textemdash{} ageing-campaign EIS data, organised by stage as \texttt{1\_BOL\_EIS/}, \texttt{2\_MOL\_EIS/} and \texttt{3\_EOL\_EIS/}; spectra are provided at 10\%, 50\% and \rev{90\%} SoC with real impedance [$\Omega$], imaginary impedance [$\Omega$] and frequency [Hz]. The underlying ageing conditions are summarised in the Methods ageing matrix (\Cref{tab:ageing_matrix}). 
\item \path{5_Documentation/Datasheet/}\textemdash{} manufacturer datasheet material and supporting documentation used to interpret the cell specifications and dataset organisation.
\end{itemize}
\par
\endgroup

\par
\section{Conclusion}\label{sec:conclusion}
This paper presents a comprehensive testing dataset from commercial 58Ah prismatic NMC cells.
The dataset is divided into three complementary parts: electrical characterization (initialization, \gls{gitt}, \gls{cc}, dynamic-cycle, \gls{hppc}, and \gls{eis} tests), an ageing campaign covering calendar and cyclic ageing, and a cell teardown of pristine and aged cells, including material and half-cell analyses. The paper also documents in detail how the test data were obtained. The methods describe the teardown and half-cell preparation procedures, the test equipment, and the characterization protocols. They also describe the design of the ageing matrix, the structure of a monthly reference performance tests, and the post-charge lithium-stripping detection. This level of detail allows users to interpret the data correctly and to reproduce or extend the experiments.
\par
To show the usability of the dataset, several example analyses are presented. For the characterization data, the \gls{gitt} experiments yield charge and discharge \glspl{ocv} of the full cell. They also capture the temperature dependence of the \gls{ocv} relative to 25°C. The \gls{eis} measurements provide impedance spectra at \gls{bol} as functions of \gls{soc} and temperature.
An example analysis of the ageing campaign shows known degradation patterns in the capacity-fade and resistance trajectories: high temperature combined with high state of charge was associated with the most severe calendar ageing, while high charging rates produced severe cyclic ageing.
The ageing campaign also incorporated reference performance tests, \gls{eis}, \gls{gitt} measurements and $dV/dt$ relaxation analysis to track capacity fade, resistance growth, impedance evolution, changes in open-circuit voltage and signatures consistent with lithium stripping.
\par
Teardown and half-cell measurements documented electrode geometry and open-circuit potentials, while microscopy and elemental mapping characterized the pristine and aged electrodes.
The aged anode exhibited a dense surface layer, whereas the aged cathode showed particle microcracking, providing electrode-level observations alongside the ageing records.
\par
Together with the accompanying processing and plotting scripts, this dataset provides a common basis for analysing cell performance and ageing and for benchmarking battery models.

\par
\section{Data and code availability}
The dataset described in this study is openly available from Zenodo at \url{https://doi.org/10.5281/zenodo.22301764} \citep{medina2026dataset}. Processing code is also available on GitHub at \url{https://github.com/robinson-medina/Analysis_Of_NMC_Dataset}.
\par
\section{CRediT authorship contribution statement}
\textbf{Robinson S. Medina}: Writing -draft, Data processing and figure generation.
\textbf{Alessio Alberto Lodge}: Writing - draft, Writing - review, Methodology, Investigation, Data collection, Data curation. 
\textbf{Feye S.J. Hoekstra}: Writing, Methodology, Data curation, Data collection, Conceptualization, Investigation, Project Administration.
\textbf{Steven Wilkins}: Editing, Supervision, Funding acquisition.
\textbf{Damian Cupid}: Writing – draft, editing document, Data collection.
\textbf{Akzhan Bekzhanov}: Writing – draft, Editing, Data collection.
\textbf{Dominik Dvorak}: Writing – draft, Editing, Data collection.
\textbf{Hansjörg Kapeller}: Reviewer and data collector.
\textbf{Gregor Kapun}: Data collection.
\textbf{Jože Moškon}: Data collection.
\textbf{Blaž Tratnik}: Data collection.
\textbf{Igor Mele}: Conceptualization, Methodology, Writing – draft, Editing.
\textbf{Tomaž Katrašnik}: Conceptualization, Methodology, Writing – review, Editing, Supervision, Funding acquisition.
\par
\section{Declaration of Competing Interest}
The authors declare that they have no known competing financial interests or personal relationships that could have appeared to
influence the work reported in this paper.
\par
\section{Acknowledgments}
This research was conducted as part of the \textbf{NEXTBMS} project \citep{nextbms2023}, funded by the European Union under the Horizon Europe program (Call Identifier: HORIZON-CL5-2022-D2-01-09, Grant Agreement No.101103898).
Views and opinions expressed are, however, those of the author(s) only and do not necessarily reflect those of the European Union or the European Commission. Neither the European Union nor the granting authority can be held responsible for them
\par
T.K. and I.M. also acknowledge financial support from the Slovenian Research Agency (research core funding No. P2-0401).
\par
\section{Declaration of generative AI and AI-assisted technologies in the manuscript preparation process}
During the preparation of this work the authors used Github Copilot and several models within to improve the readability of the article and assist with the development of the processing code. After using this tool/service, the author(s) reviewed and edited the content as needed and take full responsibility for the content of the publication.
\par
\bibliographystyle{elsarticle-harv}
\bibliography{References}

@ARTICLE{Guenther2013ModelbasedIO,
  title={Model-based investigation of electric vehicle battery aging by means of vehicle-to-grid scenario simulations},
  author={Clemens Guenther and Benjamin Schott and Wilfried Hennings and Paul Waldowski and Michael A. A Danzer},
  journal={Journal of Power Sources},
  year={2013},
  volume={239},
  pages={604-610},
  url={https://api.semanticscholar.org/CorpusID:109967096}
}

@article{ding2025,
title = {Prognosticating nonlinear degradation in lithium-ion batteries: operando pressure as an early indicator preceding other signals of capacity fade and safety risks},
journal = {Energy Storage Materials},
volume = {75},
pages = {103998},
year = {2025},
issn = {2405-8297},
doi = {https://doi.org/10.1016/j.ensm.2024.103998},
url = {https://www.sciencedirect.com/science/article/pii/S2405829724008250},
author = {Shicong Ding and Li Wang and Haifeng Dai and Xiangming He}
}

@techreport{kelly2023cradle,
  title={Cradle-to-grave lifecycle analysis of US light-duty vehicle-fuel pathways: a greenhouse gas emissions and economic assessment of current (2020) and future (2030-2035) technologies},
  author={Kelly, Jarod C and Elgowainy, Amgad and Isaac, Raphael and Ward, Jacob and Islam, Ehsan and Rousseau, Aymeric and Sutherland, Ian and Wallington, Timothy J and Alexander, Marcus and Muratori, Matteo and others},
  year={2023},
  institution={Argonne National Laboratory (ANL), Argonne, IL (United States)}
}

@article{prakash2022review,
  title={A review of battery energy storage systems for ancillary services in distribution grids: Current status, challenges and future directions},
  author={Prakash, Krishneel and Ali, Muhammad and Siddique, Md Nazrul Islam and Chand, Aneesh A and Kumar, Nallapaneni Manoj and Dong, Daoyi and Pota, Hemanshu R},
  journal={Frontiers in Energy Research},
  volume={10},
  pages={971704},
  year={2022},
  publisher={Frontiers Media SA}
}

@book{plett2015battery,
  title={Battery management systems, Volume I: Battery modeling},
  author={Plett, Gregory L},
  year={2015},
  publisher={Artech House}
}

@book{plett2024battery,
  title={Battery Management Systems, Volume III: Physics-Based Methods},
  author={Plett, Gregory L and Trimboli, M Scott},
  volume={3},
  year={2024},
  publisher={Artech House}
}

@article{borah2024synergizing,
  title={Synergizing physics and machine learning for advanced battery management},
  author={Borah, Manashita and Wang, Qiao and Moura, Scott and Sauer, Dirk Uwe and Li, Weihan},
  journal={Communications Engineering},
  volume={3},
  number={1},
  pages={134},
  year={2024},
  publisher={Nature Publishing Group UK London}
}

@article{lopetegi2024new,
  title={A new battery soc/soh/esoh estimation method using a pbm and interconnected spkfs: Part ii. soh and esoh estimation},
  author={Lopetegi, Iker and Plett, Gregory L and Trimboli, M Scott and Oca, Laura and Miguel, Eduardo and Iraola, Unai},
  journal={Journal of The Electrochemical Society},
  volume={171},
  number={3},
  pages={030518},
  year={2024},
  publisher={IOP Publishing}
}

@article{kench2024li,
  title={Li-ion battery design through microstructural optimization using generative AI},
  author={Kench, Steve and Squires, Isaac and Dahari, Amir and Planella, Ferran Brosa and Roberts, Scott A and Cooper, Samuel J},
  journal={Matter},
  volume={7},
  number={12},
  pages={4260--4269},
  year={2024},
  publisher={Elsevier}
}

@article{grover2018optimal,
  title={Optimal sizing and placement of distribution grid connected battery systems through an SOCP optimal power flow algorithm},
  author={Grover-Silva, Etta and Girard, Robin and Kariniotakis, George},
  journal={Applied Energy},
  volume={219},
  pages={385--393},
  year={2018},
  publisher={Elsevier}
}

@article{zhang2023optimal,
  title={Optimal sizing of behind-the-meter battery storage for providing profit-oriented stackable services},
  author={Zhang, Yichao and Anvari-Moghaddam, Amjad and Peyghami, Saeed and Li, Yuan and Dragi{\v{c}}evi{\'c}, Tomislav and Blaabjerg, Frede},
  journal={IEEE Transactions on Smart Grid},
  volume={15},
  number={2},
  pages={1481--1494},
  year={2023},
  publisher={IEEE}
}

@article{panahi2025fast,
  title={Fast and generalizable parameter-embedded neural operators for lithium-ion battery simulation},
  author={Panahi, Amir Ali and Luder, Daniel and Wu, Billy and Offer, Gregory and Sauer, Dirk Uwe and Li, Weihan},
  journal={Energy and AI},
  pages={100647},
  year={2025},
  publisher={Elsevier}
}

@article{li2021one,
  title={One-shot battery degradation trajectory prediction with deep learning},
  author={Li, Weihan and Sengupta, Neil and Dechent, Philipp and Howey, David and Annaswamy, Anuradha and Sauer, Dirk Uwe},
  journal={Journal of Power Sources},
  volume={506},
  pages={230024},
  year={2021},
  publisher={Elsevier}
}

@article{li2025fast,
  title={Fast data augmentation for battery degradation prediction},
  author={Li, Weihan and Samsukha, Harshvardhan and van Vlijmen, Bruis and Yan, Lisen and Greenbank, Samuel and Onori, Simona and Viswanathan, Venkat},
  journal={Energy and AI},
  volume={21},
  pages={100542},
  year={2025},
  publisher={Elsevier}
}

@inproceedings{roy2024forecasting,
  title={Forecasting Electrode State of Health (eSOH) for Managing Battery Lifetime Using Domain-Knowledge-Informed Machine Learning},
  author={Roy, Apoorva and Alghalayini, Maher and Movahedi, Hamidreza and Siegel, Jason and Harris, Stephen J and Stefanopoulou, Anna},
  booktitle={ECS Meeting Abstracts},
  number={3},
  pages={383--383},
  year={2024}
}

@article{wang2023large,
  title={Large-scale field data-based battery aging prediction driven by statistical features and machine learning},
  author={Wang, Qiushi and Wang, Zhenpo and Liu, Peng and Zhang, Lei and Sauer, Dirk Uwe and Li, Weihan},
  journal={Cell Reports Physical Science},
  volume={4},
  number={12},
  year={2023},
  publisher={Elsevier}
}

@article{wang2024unlocking,
  title={Unlocking the potential of unlabeled data: Self-supervised machine learning for battery aging diagnosis with real-world field data},
  author={Wang, Qiao and Ye, Min and Celik, Sehriban and Deng, Zhongwei and Li, Bin and Sauer, Dirk Uwe and Li, Weihan},
  journal={Journal of Energy Chemistry},
  volume={99},
  pages={681--691},
  year={2024},
  publisher={Elsevier}
}

@article{wassiliadis2023model,
  title={Model-based health-aware fast charging to mitigate the risk of lithium plating and prolong the cycle life of lithium-ion batteries in electric vehicles},
  author={Wassiliadis, Nikolaos and Kriegler, Johannes and Gamra, Kareem Abo and Lienkamp, Markus},
  journal={Journal of Power Sources},
  volume={561},
  pages={232586},
  year={2023},
  publisher={Elsevier}
}

@article{reniers2021unlocking,
  title={Unlocking extra value from grid batteries using advanced models},
  author={Reniers, Jorn M and Mulder, Grietus and Howey, David A},
  journal={Journal of power sources},
  volume={487},
  pages={229355},
  year={2021},
  publisher={Elsevier}
}

@article{brosa2022continuum,
  title={A continuum of physics-based lithium-ion battery models reviewed},
  author={Brosa Planella, Ferran and Ai, Weilong and Boyce, Adam M and Ghosh, Abir and Korotkin, Ivan and Sahu, Smita and Sulzer, Valentin and Timms, Robert and Tranter, Thomas G and Zyskin, Maxim and others},
  journal={Progress in Energy},
  volume={4},
  number={4},
  pages={042003},
  year={2022},
  publisher={IOP Publishing}
}

@article{chen2020development,
  title={Development of experimental techniques for parameterization of multi-scale lithium-ion battery models},
  author={Chen, Chang-Hui and Brosa Planella, Ferran and O’regan, Kieran and Gastol, Dominika and Widanage, W Dhammika and Kendrick, Emma},
  journal={Journal of The Electrochemical Society},
  volume={167},
  number={8},
  pages={080534},
  year={2020},
  publisher={IOP Publishing}
}

@article{sheikh2026pybatteryid,
  title={PyBatteryID--A Python Package for Data-Driven Battery Model Identification using LPV Framework},
  author={Sheikh, Muiz and Donkers, Tijs and Bergveld, Henk Jan},
  journal={Journal of Open Research Software},
  volume={14},
  number={1},
  year={2026}
}

@article{hoekstra2023rapid,
  title={Rapid empirical battery electromotive-force and overpotential modelling using input--output linear parameter-varying methods},
  author={Hoekstra, FSJ and Donkers, MCF and Bergveld, HJ},
  journal={Journal of Energy Storage},
  volume={65},
  pages={107185},
  year={2023},
  publisher={Elsevier}
}

@misc{aboutenergy,
  author       = {{About:Energy Ltd.}},
  title        = {{About:Energy} -- Battery Pack Design \& Cell Modelling Tools},
  howpublished = {\url{https://www.aboutenergy.io/}},
  year         = {2026},
  note         = {Accessed: 2026-05-11}
}

@misc{batterydesign,
  author       = {Taylor, Nigel},
  title        = {{BatteryDesign.net} -- Designed by Battery Engineers for Battery Engineers},
  howpublished = {\url{https://www.batterydesign.net/}},
  year         = {2026},
  note         = {Accessed: 2026-05-11}
}

@misc{batemo,
  author       = {{Batemo GmbH}},
  title        = {Batemo -- Battery Modelling and Battery Simulation Solutions},
  howpublished = {\url{https://www.batemo.com/}},
  year         = {2026},
  note         = {Accessed: 2026-05-11}
}

@article{catenaro2021experimental,
  title={Experimental data of lithium-ion batteries under galvanostatic discharge tests at different rates and temperatures of operation},
  author={Catenaro, Edoardo and Onori, Simona},
  journal={Data in Brief},
  volume={35},
  pages={106894},
  year={2021},
  publisher={Elsevier}
}

@article{khan2025high,
  title={High-power lithium-ion battery characterization dataset for stochastic battery modeling},
  author={Khan, Muhammad Aadil and Thatipamula, Sai and Tresca, Luigi and Xu, Le and Trewartha, Amalie and Onori, Simona},
  journal={Scientific Data},
  volume={12},
  number={1},
  pages={1506},
  year={2025},
  publisher={Nature Publishing Group UK London}
}

@article{fasolato2025dataset,
  title={A dataset for large prismatic lithium-ion battery cells (CALB L148N58A): Comprehensive characterization and real-world driving cycles},
  author={Fasolato, Simone and Blasuttigh, Nicola and Galuppini, Giacomo and Raimondo, Davide M},
  journal={Data in Brief},
  volume={58},
  pages={111301},
  year={2025},
  publisher={Elsevier}
}

@article{kirkaldy2024lithium,
  title={Lithium-ion battery degradation: Comprehensive cycle ageing data and analysis for commercial 21700 cells},
  author={Kirkaldy, Niall and Samieian, Mohammad A and Offer, Gregory J and Marinescu, Monica and Patel, Yatish},
  journal={Journal of Power Sources},
  volume={603},
  pages={234185},
  year={2024},
  publisher={Elsevier}
}

@article{stroebl2024multi,
  title={A multi-stage lithium-ion battery aging dataset using various experimental design methodologies},
  author={Stroebl, Florian and Petersohn, Ronny and Schricker, Barbara and Schaeufl, Florian and Bohlen, Oliver and Palm, Herbert},
  journal={Scientific data},
  volume={11},
  number={1},
  pages={1020},
  year={2024},
  publisher={Nature Publishing Group UK London}
}

@article{luh2024comprehensive,
  title={Comprehensive battery aging dataset: capacity and impedance fade measurements of a lithium-ion NMC/C-SiO cell},
  author={Luh, Matthias and Blank, Thomas},
  journal={Scientific Data},
  volume={11},
  number={1},
  pages={1004},
  year={2024},
  publisher={Nature Publishing Group UK London}
}

@article{khan2025calendar,
  title={Calendar and cycle aging dataset for lithium-ion batteries with multi-SOC electrochemical impedance spectroscopy measurements},
  author={Khan, Muhammad Aadil and Chu, Colin and Onori, Simona},
  journal={Data in Brief},
  pages={112282},
  year={2025},
  publisher={Elsevier}
}

@article{pozzato2022lithium,
  title={Lithium-ion battery aging dataset based on electric vehicle real-driving profiles},
  author={Pozzato, Gabriele and Allam, Anirudh and Onori, Simona},
  journal={Data in brief},
  volume={41},
  pages={107995},
  year={2022},
  publisher={Elsevier}
}

@article{moy2024second,
  title={Second-life lithium-ion battery aging dataset based on grid storage cycling},
  author={Moy, Kevin and Khan, Muhammad Aadil and Fasolato, Simone and Pozzato, Gabriele and Allam, Anirudh and Onori, Simona},
  journal={Data in Brief},
  volume={57},
  pages={111046},
  year={2024},
  publisher={Elsevier}
}

@article{Cui_2025,
doi = {10.1149/1945-7111/adde16},
url = {https://doi.org/10.1149/1945-7111/adde16},
year = {2025},
month = {jun},
publisher = {IOP Publishing},
volume = {172},
number = {6},
pages = {060521},
author = {Cui, Xiaofan and Stroebl, Florian and Uppaluri, Maitri and Lam, Vivek N. and Chueh, William C. and Onori, Simona},
title = {Long-Term Calendar Aging Across Commercial Lithium-Ion Cell Chemistries—Modeling and Early Prediction},
journal = {Journal of The Electrochemical Society}
}

@article{ank2023lithium,
  title={Lithium-ion cells in automotive applications: Tesla 4680 cylindrical cell teardown and characterization},
  author={Ank, Manuel and Sommer, Alessandro and Abo Gamra, Kareem and Sch{\"o}berl, Jan and Leeb, Matthias and Schachtl, Johannes and Streidel, Noah and Stock, Sandro and Schreiber, Markus and Bilfinger, Philip and others},
  journal={Journal of The Electrochemical Society},
  volume={170},
  number={12},
  pages={120536},
  year={2023},
  publisher={IOP Publishing}
}

@article{gunter2022state,
  title={State of the art of lithium-ion pouch cells in automotive applications: cell teardown and characterization},
  author={G{\"u}nter, FJ and Wassiliadis, Nikolaos},
  journal={Journal of The Electrochemical Society},
  volume={169},
  number={3},
  pages={030515},
  year={2022},
  publisher={IOP Publishing}
}

@article{stock2023cell,
  title={Cell teardown and characterization of an automotive prismatic LFP battery},
  author={Stock, Sandro and Hagemeister, Jan and Grabmann, Sophie and Kriegler, Johannes and Keilhofer, Josef and Ank, Manuel and Dickmanns, Jonas LS and Schreiber, Markus and Konwitschny, Fabian and Wassiliadis, Nikolaos and others},
  journal={Electrochimica Acta},
  volume={471},
  pages={143341},
  year={2023},
  publisher={Elsevier}
}

@misc{nextbms2023,
  title        = {{NEXT-generation physics and data-based Battery Management Systems for optimised battery utilization (NEXTBMS)}},
  author       = {{European Commission}},
  year         = {2023},
  note         = {Horizon Europe, Grant Agreement No. 101103898. Project website: \url{https://nextbms.eu/}},
  howpublished = {CORDIS},
  doi          = {10.3030/101103898},
  url          = {https://doi.org/10.3030/101103898}
}

@techreport{iea2024batteries,
  author      = {{IEA}},
  title       = {Batteries and Secure Energy Transitions},
  institution = {IEA},
  year        = {2024},
  address     = {Paris},
  url         = {https://www.iea.org/reports/batteries-and-secure-energy-transitions},
  note        = {Licence: CC BY 4.0}
}

@techreport{iea2026globalenergyreview,
  author      = {{IEA}},
  title       = {Global Energy Review 2026},
  institution = {IEA},
  year        = {2026},
  address     = {Paris},
  url         = {https://www.iea.org/reports/global-energy-review-2026},
  note        = {Licence: CC BY 4.0}
}

@article{SAEED2024102436,
title = {Calendar aging of commercial lithium ion batteries under realistic conditions},
journal = {Thermal Science and Engineering Progress},
volume = {49},
pages = {102436},
year = {2024},
issn = {2451-9049},
doi = {https://doi.org/10.1016/j.tsep.2024.102436},
url = {https://www.sciencedirect.com/science/article/pii/S2451904924000544},
author = {Ahmed Saeed and Ali Alshehri and Obaidallah Munteshari}
}

@article{SMITH2023233118,
title = {Localized lithium plating under mild cycling conditions in high-energy lithium-ion batteries},
journal = {Journal of Power Sources},
volume = {573},
pages = {233118},
year = {2023},
issn = {0378-7753},
doi = {https://doi.org/10.1016/j.jpowsour.2023.233118},
url = {https://www.sciencedirect.com/science/article/pii/S0378775323004937},
author = {Alexander J. Smith and Yuan Fang and Anastasiia Mikheenkova and Henrik Ekström and Pontus Svens and Istaq Ahmed and Matthew J. Lacey and Göran Lindbergh and István Furó and Rakel Wreland Lindström}
}

@article{frie2024analysis,
  title={An Analysis of Calendaric Aging over 5 Years of Ni-rich 18650-Cells with Si/C Anodes},
  author={Frie, Fabian and Ditler, Heinrich and Klick, Sebastian and Stahl, Gereon and Rahe, Christiane and Ghaddar, Tala and Sauer, Dirk Uwe},
  journal={ChemElectroChem},
  volume={11},
  number={9},
  pages={e202400020},
  year={2024},
  publisher={Wiley Online Library}
}

@article{SCHMITT2017183,
title = {Impedance change and capacity fade of lithium nickel manganese cobalt oxide-based batteries during calendar aging},
journal = {Journal of Power Sources},
volume = {353},
pages = {183-194},
year = {2017},
issn = {0378-7753},
doi = {https://doi.org/10.1016/j.jpowsour.2017.03.090},
url = {https://www.sciencedirect.com/science/article/pii/S037877531730397X},
author = {Julius Schmitt and Arpit Maheshwari and Michael Heck and Stephan Lux and Matthias Vetter},
}

@article{klick2025failure,
  title={Failure Mode and Degradation Analysis of a Commercial Sodium-Ion Battery With Severe Gassing Issue},
  author={Klick, Sebastian and Laufen, Hendrik and Sch{\"u}tte, Moritz and Qian, Bowen and Quade, Katharina Lilith and Rahe, Christiane and Dubarry, Matthieu and Sauer, Dirk Uwe},
  journal={Batteries \& Supercaps},
  volume={8},
  number={4},
  pages={e202400546},
  year={2025},
  publisher={Wiley Online Library}
}

@article{auch2026new,
  title={New Insights Into Commercial 18650 Sodium-Ion Batteries: Understanding Cell Characteristics and Ageing Through Multi Method Analyses},
  author={Auch, Bernhard and Weisenberger, Christian and Kaya, Pinar and Knoblauch, Volker},
  journal={Batteries \& Supercaps},
  volume={9},
  number={3},
  pages={e202500925},
  year={2026},
  publisher={Wiley Online Library}
}

@article{he2023high,
  title={High power NVPF/HC-based sodium-ion batteries},
  author={He, Minglong and Mejdoubi, Asmae EL and Chartouni, Daniel and Morcrette, Mathieu and Troendle, Pirmin and Castiglioni, Roberto},
  journal={Journal of Power Sources},
  volume={588},
  pages={233741},
  year={2023},
  publisher={Elsevier}
}

@article{streck2024self,
  title={Self-discharge and calendar aging behavior of Li-ion and Na-ion cells},
  author={Streck, Luiza and Roth, Thomas and Bosch, Hannah and Kirst, Cedric and Rehm, Mathias and Keil, Peter and Jossen, Andreas},
  journal={Journal of The Electrochemical Society},
  volume={171},
  number={8},
  pages={080531},
  year={2024},
  publisher={IOP Publishing}
}

@article{kassem2012calendar,
  title={Calendar aging of a graphite/LiFePO4 cell},
  author={Kassem, Mohammad and Bernard, Julien and Revel, Renaud and Pelissier, Serge and Duclaud, Fran{\c{c}}ois and Delacourt, Charles},
  journal={Journal of Power Sources},
  volume={208},
  pages={296--305},
  year={2012},
  publisher={Elsevier}
}

@article{laufen2024multi,
  title={Multi-method characterization of a commercial 1.2 Ah sodium-ion battery cell indicates drop-in potential},
  author={Laufen, Hendrik and Klick, Sebastian and Ditler, Heinrich and Quade, Katharina Lilith and Mikitisin, Adrian and Bloemeke, Alexander and Schuette, Moritz and Wasylowski, David and Sonnet, Morian and Henrich, Laura and others},
  journal={Cell Reports Physical Science},
  volume={5},
  number={5},
  year={2024},
  publisher={Elsevier}
}

@article{chahbaz2021non,
  title={Non-invasive identification of calendar and cyclic ageing mechanisms for lithium-titanate-oxide batteries},
  author={Chahbaz, Ahmed and Meishner, Fabian and Li, Weihan and {\"U}nl{\"u}bayir, Cem and Sauer, Dirk Uwe},
  journal={Energy storage materials},
  volume={42},
  pages={794--805},
  year={2021},
  publisher={Elsevier}
}

@article{konz2020detecting,
  title={Detecting the onset of lithium plating and monitoring fast charging performance with voltage relaxation},
  author={Konz, Zachary M and McShane, Eric J and McCloskey, Bryan D},
  journal={ACS Energy Letters},
  volume={5},
  number={6},
  pages={1750--1757},
  year={2020},
  doi={10.1021/acsenergylett.0c00831},
  publisher={ACS Publications}
}

@misc{medina2026dataset,
  author       = {Medina, Robinson and Hoekstra, Feye and Lodge, Alessio Alberto and Wilkins, Steven and Cupid, Damian and Bekzhanov, Akzhan and Dvorak, Dominik and Kapeller, Hansj{\"o}rg and Mo{\v{s}}kon, Jo{\v{z}}e and Mele, Igor and Katra{\v{s}}nik, Toma{\v{z}}},
  title        = {Characterization, ageing and post-mortem dataset of commercial 58{Ah} prismatic {NMC} cells},
  howpublished = {Zenodo, dataset},
  year         = {2026},
  doi          = {10.5281/zenodo.22301764},
  url          = {https://doi.org/10.5281/zenodo.22301764}
}

@article{rashid2015relaxation,
  author  = {Rashid, Muhammad and Gupta, Amit},
  title   = {Effect of Relaxation Periods over Cycling Performance of a {Li}-Ion Battery},
  journal = {Journal of The Electrochemical Society},
  volume  = {162},
  number  = {2},
  pages   = {A3145--A3153},
  year    = {2015},
  doi     = {10.1149/2.0201502jes},
}

@article{schreiber2025relaxation,
  author  = {Schreiber, Markus and Steiner, Theresa and Kayl, Jonas and Sch{\"o}nberger, Benedikt and Grosu, Cristina and Lienkamp, Markus},
  title   = {The Overlooked Role of Battery Cell Relaxation: How Reversible Effects Manipulate Accelerated Aging Characterization},
  journal = {World Electric Vehicle Journal},
  volume  = {16},
  number  = {5},
  pages   = {255},
  year    = {2025},
  doi     = {10.3390/wevj16050255},
}
\balance
\end{document}